\documentclass[12pt]{article}
\usepackage{amsmath}
\usepackage{graphicx}
\usepackage{natbib}

\newcommand{\blind}{0}

\usepackage{amssymb}
\usepackage{amsmath}
\usepackage{amsthm}
\usepackage{hyperref,bbm}
\usepackage{amsfonts}
\usepackage{amsbsy}
\usepackage{color}
\usepackage{enumitem}
\usepackage{listings}
\usepackage{booktabs}
\usepackage{algpseudocode}
\usepackage{algorithm}
\usepackage{xr-hyper}
\usepackage{tcolorbox}
\usepackage{xcolor}
\usepackage{empheq}

\usepackage[title]{appendix}

\DeclareMathOperator*{\argmin}{argmin}
\definecolor{slateblue}{RGB}{131, 111, 255}
\definecolor{deeppink}{RGB}{255, 20, 147}

\def\tcr{\textcolor{black}}

\newcommand{\tr}{\mathrm{tr}}
\newcommand{\sign}{\mathrm{sign}}
\newcommand{\diag}{\mathrm{diag}}
\newcommand{\vc}{\mathrm{vec}}

\newcommand{\vh}{\mathrm{vech}}

\newcommand{\tp}{\mathsf{T}}
\newcommand{\grad}[1]{\frac{\partial \ell_n(#1)}{\partial\theta}}

\newcommand{\gradEc}[1]{\frac{\partial \ell_n(#1)}{\partial\theta_{E'}}}

\newtheorem{definition}{Definition}
\newtheorem{theorem}{Theorem}

\newtheorem{proposition}{Proposition}
\newtheorem{lemma}{Lemma}

\newcommand{\graybox}[1]{%
  \fcolorbox{black}{gray!10}{#1}%
}

\newtcolorbox{grayenum}{
  colback=gray!10,
  colframe=gray!60,
  boxrule=0.8pt,
  arc=0pt,        % square corners (like empheq)
  left=6pt,
  right=6pt,
  top=6pt,
  bottom=6pt
}

\begin{document}

	\def\spacingset#1{\renewcommand{\baselinestretch}%
		{#1}\small\normalsize} \spacingset{1}

	%%%%%%%%%%%%%%%%%%%%%%%%%%%%%%%%%%%%%%%%%%%%%%%%%%%%%%%%%%%%%%%%%%%%%%%%%%%%%%
	
	\if0\blind
	{
		\title{\bf Selective Inference in Graphical Models via Maximum Likelihood}
		\author{Sofia Guglielmini\thanks{
				The authors acknowledge support from project C16/20/002 of the Research Fund KU Leuven, Belgium, project ASTeRISK Research Foundation Flanders [grant number 40007517] of the Excellence of Science (EOS) program (FWO-FNRS, Belgium) and NSF CAREER Award DMS-2337882. Data are provided by OASIS (OASIS-1). Principal Investigators: D. Marcus, R, Buckner, J, Csernansky J. Morris; P50 AG05681, P01 AG03991, P01 AG026276, R01 AG021910, P20 MH071616, U24 RR021382.}\hspace{.2cm}\\
			Orstat and Leuven Statistics Research Center, KU Leuven, Leuven, Belgium,\\
			Gerda Claeskens \\
			Orstat and Leuven Statistics Research Center, KU Leuven, Leuven, Belgium \\
			and \\
			Snigdha Panigrahi \\
			Department of Statistics, University of Michigan \\}
		\date{}
		\maketitle
	} \fi
	
	\if1\blind
	{
		\bigskip
		\bigskip
		\bigskip
		\begin{center}
			{\LARGE\bf Selective Inference in Graphical Models via Maximum Likelihood}
		\end{center}
		\medskip
	} \fi
	
	\bigskip
	\begin{abstract}
	The graphical lasso is a widely used algorithm for fitting undirected Gaussian graphical models.
However, for inference on functionals of edge values in the learned graph, standard tools lack formal statistical guarantees, such as control of the type I error rate.
We introduce a selective inference method for asymptotically valid inference after graphical lasso selection with added randomization.
We obtain a selective likelihood, conditional on the event of selection, through a change of variable on the known density of the randomization variables.
Our method enables interval estimation and hypothesis testing for a wide range of functionals of edge values in the learned graph using the conditional maximum likelihood estimator.
Our numerical studies show that introducing a small amount of randomization: (i) achieves better results than data splitting in terms of model selection quality; (ii) ensures intervals of bounded length also in high-dimensional settings where data splitting suffers from low power or does not reach asymptotic regime due to limited samples for inference; (iii) enables inference for a wide range of inferential targets in the learned graph, including partial correlations and measures of node influence and connectivity between nodes.
       \end{abstract}
	
	\noindent%
	{\it Keywords:}  selective inference, randomization, network analysis, graphical models
	\vfill
	
	\newpage
	\spacingset{1.75} % DON'T change the spacing!

\section{Introduction}

The graphical lasso is a popular algorithm for learning sparse, undirected Gaussian graphical models.
The model assumes that the data consist of $n$ independent and identically distributed observations $X=(X_1,\ldots,X_n)^\tp$, drawn from a centered $p$-dimensional normal distribution:
%\begin{equation*}
$
	X_h \sim N_p(0_p, \Theta^{-1})$,
	% \quad \text{ for all } \hspace{1mm} 
	for $h = \{1,\ldots,n\}$.
%\end{equation*}
The presence or absence of edges in the graph underlying this model, where the nodes represent variables, is encoded in the precision matrix $\Theta$, the inverse of the covariance matrix. 
More specifically, the edge between nodes $i$ and $j$ is absent if and only if the $(i,j)$-th element of the precision matrix $\Theta$ is equal to 0; equivalently, by the pairwise Markov property, if and only if $X_i$ and $X_j$ are independent conditionally on the other variables in the graph.

To encourage sparsity in the edges of the graph, the graphical lasso imposes a penalty on the $\ell_1$-norm of the precision matrix $\Theta$.
Given a sample covariance matrix $C_n=X^\tp X/n$ and a positive penalty parameter $\lambda$ with asymptotic order $O(n^{-1/2})$, the graphical lasso estimator is obtained as the solution of the convex problem \citep{yuan2007model}
\begin{equation}\label{chapter3:lasso}
	\argmin_{\substack{\hspace{2.5mm}\{\Theta=\Theta^\tp,\;\Theta\succ0\}}} \left\{\tr(C_n\Theta) -\log\det(\Theta)+  \lambda ||\Theta||_1\right\},
\end{equation}
\tcr{where $||\Theta||_1 = \sum_{i,j}|\Theta_{i,j}|$.}
% is the entrywise $\ell_1$ norm of $\Theta$
The loss part of the graphical lasso problem 
is proportional to the multivariate Gaussian negative log-likelihood,
\begin{equation}\label{chapter3:likelihood}
	\ell_n(\Theta;X) = \frac{n}{2}\tr(C_n\Theta)-\frac{n}{2}\log\det(\Theta).
\end{equation}
The optimization is carried over the class of $p\times p$ symmetric matrices that are positive definite, as denoted by $\Theta\succ0$, and can be efficiently solved using fast iterative algorithms, such as the method in \citet{friedman2008sparse}.

In many cases, learning a graph is not the final step of the analysis.
From a scientific perspective, inference is of primary importance. Inference can consist of testing the significance of the edges, quantifying uncertainty in the estimates of edge values or constructing confidence intervals.
More generally, one may seek to answer questions about functionals of edge parameters in the learned graph.
In network analysis, practical questions may involve estimating partial correlations, measuring the centrality of a node, or estimating connectivity measures between groups of nodes in the learned graph.
Such questions involve ``post-selection'' parameters that are functions of the learned graph or the entries of the estimated precision matrix.
These questions cannot be addressed using naive inferential tools, which use the same data to first select parameters and then seek inferential answers.

Modern selective inference methods address this challenge by conditioning on the outcome of the selection procedure, i.e., the event of selection.
In this paper, we introduce a conditional selective inference method for edge parameters in graphs estimated using the graphical lasso.
Building on the likelihood-based framework in \citet{panigrahi2023approximate}, we develop a maximum likelihood approach for graphical models. 

\tcr{Central to our approach is the randomization in the selection. The selected graph depends on the observed data and on an auxiliary randomization variable with known distribution. The first-order optimality conditions of the randomized graphical lasso induce a map between the unknown distribution of the penalized estimators and the known distribution of the randomization variable. A change-of-variable argument based on this map yields the selective likelihood, conditional on the selection event. At the same time, randomization reduces the amount of information that is used in selection, preserving more for inference and thus resulting in improved inferential power.}

\subsection{Related work and contributions}

\tcr{Data splitting avoids reusing the data by using different samples for selection and inference.
By discarding samples in both steps, it yields worse selection and reduced power in inference, often resulting in inadmissible tests \citep{fithian2017optimal}.}
\tcr{Selective inference approaches without randomization include the polyhedral method in \citet{lee2016exact} and its extension to graphical models \citep{taylor2018post, guglielmini2025asymptotic}.}

The advantages of our maximum likelihood method compared to existing selective inference approaches are summarized below.
\begin{enumerate}[leftmargin=*]
	\item[(i)] Randomization in selection allows the \emph{full} dataset to be used for \tcr{selection and} inference while preserving type I error rate control. With a small amount of randomization, the resulting selection accuracy is comparable to that obtained by allocating a large proportion of the data to the selection step in data splitting. 
	%\tcr{Unlike data splitting, our method retains the full dataset for inference, allowing the inferential procedure to benefit from the full sample size and reach its asymptotic regime.}

	\item[(ii)] Our method yields substantially more powerful inference compared to conditional polyhedral methods \citep{lee2016exact,taylor2018post,guglielmini2025asymptotic}.
	
	\item[(iii)] Our framework enables inference for general functionals of edge parameters, including quantities used in network analysis such as partial correlations, connectivity, and node strength. Standard tools such as the bootstrap or delta method can be applied to obtain confidence intervals and hypothesis tests.
	In contrast, these tasks \emph{cannot} generally be achieved with earlier conditional approaches that form a pivot or a p-value for a scalar-valued parameter. This includes %Such approaches have been developed for selective inference in different contexts, such as 
\citet{lee2016exact, tian2018selective, panigrahi2018scalable, duy2021parametric, panigrahi2024exact} for lasso regression, \citet{huang2025selectivegraph} for the neighborhood lasso approach to learn graphs, \citet{neufeld2022tree, bakshi2024inference} for regression trees, \citet{freidling2024selective} for adaptive experiments.
\end{enumerate}

Finally, we emphasize that our method is designed to address a different inferential objective compared to the debiased approach for high-dimensional graphical lasso in \citet{jankova2015confidence}. 
Such methods provide confidence intervals for all entries of the precision matrix but do not enable estimation or hypothesis testing for edge parameters in a learned graph and thus do not exploit the learned structure.

The rest of the paper is organized as follows.
In Section~\ref{chapter3:sec:randomized}, we introduce the randomized graphical lasso with additive noise. Section~\ref{chapter3:sec:targets} contains various interesting post-selection inference quantities. 
% that practitioners seek to estimate in the learned graph.
Section~\ref{chapter3:sec:framework} defines the selection event underlying the graph estimated by the graphical lasso algorithm and derives the corresponding selective likelihood. Using this likelihood, we obtain the conditional maximum likelihood estimator (MLE) and explain how to compute confidence intervals and hypothesis tests.
Finally, numerical experiments in Section~\ref{chapter3:sec:simulations} and a data application in Section~\ref{chapter3:sec:application} demonstrate our method's performance across different types of graph data, highlighting its advantages over existing approaches in terms of coverage control, inferential power and applicability.
The proofs are collected in the Appendix.

\section{The randomized graphical lasso}\label{chapter3:sec:randomized}

Define $d = p(p+1)/2$ and denote by $\theta=\vh(\Theta) \in \mathbb{R}^d$ the vector-half of the symmetric matrix $\Theta\in\mathbb{R}^{p\times p}$, which collects the entries $\{\Theta_{i,j}: i \leq j\}$.
\tcr{Throughout this paper, we consider the asymptotic regime where $n \to \infty$ and $p$ is fixed.}

\tcr{To introduce random noise in the selection step, we use a Gaussian randomization variable. Besides being the standard choice in randomized selective inference \citep{huang2025selective, panigrahi2024exact}, it considerably simplifies the derivation of the likelihood presented in Section~\ref{chapter3:sec:inference} and shares an asymptotically equivalence with sample splitting \cite{panigrahi2021integrative}. Alternative choices, such as logistic and Laplace, have been considered in \citet{tian2018selective}.}
Generate a random variable $\omega_n=\omega/\sqrt{n}$, where $\omega \sim N_d(0_d, \Omega)$ with a fixed, pre-specified $\Omega$. Take then the $p\times p$ matrix $W_n$ such that $D_p^\tp\vc(W_n)/2=\omega_n$, \tcr{where $\vc(W_n)$ is the vectorization of $W_n$} and $D_p$ is the duplication matrix such that $\vc(A) = D_p \vh(A)$. In practice, $W_n$ corresponds to the symmetric matrix with vector-half equal to $\omega_n$, with the diagonal multiplied by 2 \citep[][Theorem 3.14(a)]{magnus2019matrix}.

The proposed randomized graphical lasso estimator $\widehat\Theta$ is 
defined as
%the solution of the optimization problem
\begin{equation} \label{chapter3:Theta_rho}
	\widehat\Theta = \argmin_{\substack{\hspace{2.5mm}\{\Theta=\Theta^\tp, \ \Theta\succ0}\}} \left\{\tr((C_n-W_n)\Theta)-\log\det(\Theta) +\lambda||\Theta||_1\right\}, 
\end{equation}
\tcr{where $\lambda>0$ is a penalty parameter with asymptotic order $O(n^{-1/2})$.}

\begin{proposition}[Strict convexity of the randomized graphical lasso]\label{chapter3:rglasso_convexity}
	Let $C_n, W_n \in \mathbb{R}^{p\times p}$ be symmetric matrices. For $\lambda>0$, the randomized graphical lasso objective function of \eqref{chapter3:Theta_rho} is \tcr{strictly convex} over the cone of symmetric positive definite matrices. Consequently, if a minimizer exists, it is unique.
\end{proposition}
%As in the graphical lasso problem in Equation~\eqref{chapter3:lasso}, 
The positive-definiteness of the solution is enforced by the $\log\det(\Theta)$ barrier function.
\tcr{However, the finite-sample existence of a minimizer is not guaranteed, since $C_n-W_n$ might not be positive semidefinite, preventing using standard arguments for the existence of a solution \citep{carter2025existence}.
Nevertheless, in all our simulations, the optimization algorithm converged to a solution. Convergence was verified by checking the explicit Karush-Kuhn-Tucker (K.K.T.) conditions.
In Section~\ref{chapter3:sec:elasticnet}, we discuss an extension to the elastic net penalty, which guarantees the existence of a solution.
}

Consistent with our notational scheme, let $\widehat\theta = \vh(\widehat\Theta)$.
We denote the set of active entries of the penalized solution by
$\widehat E = \left\{ j \in \{1,\ldots,d\} : \widehat\theta_j \neq 0\right\}$
and denote by $E$ and $E'$ the realized value of $\widehat E$ and the complement of $\widehat E$, respectively.
We thus define the selected model as the set
\begin{equation}
	\label{chapter3:RE}
	\mathcal{R}_E = \left\{ \Theta \in \mathbb{R}^{p\times p}: [\vh(\Theta)]_{E'} = 0_{|E'|}, \Theta^\tp=\Theta  \text{ and } \Theta\succ0\right\}.
\end{equation}
The set $\widehat E$ depends on the data. This data dependence creates the challenge that our paper addresses and is the reason the proposed method is necessary.

\section{Post-selection targets}\label{chapter3:sec:targets}

Once a graph is learned, and a value $E$ is observed for the set of selected edges $\widehat E$, a key challenge is constructing interval estimates for various inferential targets.
In post-selection inference, the target of inference is the parameter of interest within the selected model $\mathcal{R}_E$. 
%This value, also called the \textit{pseudo-true value}, is the closest value to the true one within the selected model. Since the selected model, or equivalently the set 
Since $\mathcal{R}_E$ depends on the data, the target maintains this data dependence through $E$.
Taking the expectation with respect to the distribution of $X$,
%With $\ell_n$ %the loss being minimized 
%the negative log-likelihood in Equation~\eqref{chapter3:likelihood},
 define the pseudo-true value as
\begin{equation}\label{chapter3:pseudotrue}
	\Theta^*_E =
	\underset{\Theta \in \mathcal{R}_E}{\argmin}\hspace{1mm} \mathbb{E}\left[\ell_n(\Theta;X)\right].
\end{equation}

Our framework provides inference for a broad class of functionals of the edge parameters in the learned graph. These include partial correlations, centrality and connectivity measures widely used in network analysis, such as node strength \citep{freeman1978centrality,barrat2004architecture}, expected influence
\citep{robinaugh2016identifying} and bridge centrality \citep{castro2019differential}. Table~\ref{tab:targets} defines the targets and lists the inference procedures.
The selective maximum likelihood estimator forms the basis of our approach; the selective delta method and selective bootstrap extend inference to nonlinear functionals. Algorithm~\ref{chapter3:algorithm1} describes the computation of the selective MLE on the selected support set~$E$.

\section{Proposed inferential framework}\label{chapter3:sec:framework}

We formalize the proposed framework for conducting inference after model selection via the randomized graphical lasso. This requires characterizing the selection event, deriving the corresponding conditional distribution and obtaining a new selective MLE and its asymptotic distribution. We further outline how inference can be conducted for functionals of the selected edges.

\subsection{\texorpdfstring{The randomized graphical lasso selection event $\{\widehat E = E\}$}{The randomized graphical lasso selection event}}

%We characterize the selection event $\{\widehat E = E\}$ in terms of the randomized graphical lasso estimator obtained by solving \eqref{chapter3:Theta_rho}.

Define the following selection-related quantities, where $\widehat S$ and $\widehat U$ represent the gradient and subgradient of the penalty evaluated at the solution of \eqref{chapter3:Theta_rho}
\begin{equation*}
	\widehat V = |\widehat\theta_E|, \quad \widehat S = \sign(\widehat\theta_E); \quad\text{and}\quad \widehat U \in [-1,1]^{|E'|}.
\end{equation*}

In order to obtain a tractable conditional distribution, take a proper subset of the selection event with $S$ and $U$ the realized values of $\widehat S$ and $\widehat U$ and \tcr{$\mathbb{R}_+^{|E|}$ the set of vectors in $\mathbb{R}^{|E|}$ with strictly positive components}:
\begin{equation}\label{chapter3:event}
	\left\{\widehat E = E, \widehat S = S, \widehat U=U\right\} = \left\{\sqrt{n}\widehat V \in \mathbb{R}_+^{|E|}, \widehat S = S, \widehat U=U\right\}.
\end{equation}
The equality holds because conditioning on the edge parameters in $E$ being selected is equivalent to conditioning on the event that their absolute values $\widehat V$ are strictly positive. 
This is analogous to the least-squares case \citep{panigrahi2024exact}.
%Conditioning on this selection event is analogous to conditioning on the selection and on the subgradient of the lasso penalty, as in the least-squares case \citep{panigrahi2024exact}.
Conditioning on a proper subset of the selection event always guarantees valid selective inference.
Even with this additional conditioning, it is important to note that, unlike data splitting, we do not condition on the entire dataset used for selection.
Consequently, as our simulations demonstrate, at similar levels of selection accuracy, our tests are more powerful than those obtained through the data splitting approach.

\subsection{The randomization map}

\tcr{The key idea to derive the selective likelihood in Section~\ref{chapter3:sec:inference} is to map the selected-model quantities $(\widehat V$, $\widehat U)$ with unknown distribution to the known randomization variable $\omega_n$. }
%\tcr{In order to derive the selective likelihood described in Section~\ref{chapter3:sec:inference}, we require the distribution of the estimator $\bar\theta_E$, conditional on the selection event, which is determined by $(\widehat V$, $\widehat U)$. The joint distribution of these quantities is not known, thus the conditional distribution is not available. However, the distribution of the randomization variable $\omega_n$ is fully known. The key idea is therefore to derive a map that links the selected-model quantities to the known randomization variable.}
%Starting from the optimization problem in \eqref{chapter3:Theta_rho}, define a map which, given $E$ and $S$, relates the randomized graphical lasso estimators $(\widehat V, \widehat U)$ to the randomization variable $\omega_n$. 
The construction of the map in Proposition~\ref{chapter3:map} relies on several auxiliary quantities.
Firstly, we define the second empirical moment of the gradient and empirical Hessian of the loss function $\ell_n(X;\Theta)$. Introducing the notation $\ell(\Theta;X_h)$ for the contribution to the negative log-likelihood of a single observation $X_h$, define the $d\times d$ matrices
\begin{align*}
	H_n(\Theta^*_E) = \frac{1}{n}\sum_{h=1}^n\frac{\partial^2 \ell(\Theta^*_E;X_h)}{\partial\theta\partial\theta^\tp}, J_n(\Theta^*_E) = \frac{1}{n}\sum_{h=1}^n\left[\frac{\partial \ell(\Theta^*_E;X_h)}{\partial\theta}\frac{\partial \ell(\Theta^*_E;X_h)}{\partial\theta}^\tp\right].
\end{align*}
Naturally, $\Theta^*_E$ is unknown, but in practice it can be substituted by any consistent estimator.
Without loss of generality, we regroup the entries of any $d\times d$ matrix $M$, such that the row and column indices for a given set $E$ appear in the top left submatrix. We define the following partitioning
$M= \begin{pmatrix}
	M_{EE} & M_{EE'} \\
	M_{E'E} & M_{E'E'}
\end{pmatrix}
= \begin{pmatrix}
	M_{E} & M_{E'}
\end{pmatrix}.
$
We do not further denote the dependence of the matrices $H_n(\Theta^*_E)$ and $J_n(\Theta^*_E)$ and their submatrices on $\Theta^*_E$.
\begin{lemma}\label{chapter3:hj_posdef}
	For any positive definite $\Theta$, the matrix $H_{n,EE}(\Theta)$ is positive definite.
	Furthermore, \tcr{if $J_{n,EE}(\Theta)$ has full rank,} then it is positive definite.
\end{lemma}
If $n<|E|$, $\text{rank}(J_{n,EE}) \le n < |E|$, so $J_{n,EE}$ is singular. In this case, substitute $J_n$ by $H_n$, which, by Lemma~\ref{chapter3:hj_posdef}, is invertible. For Gaussian data, these matrices are asymptotically equivalent by the second Bartlett identity.

We now introduce the \textit{naive estimator}, denoted by $\bar\Theta_E$ and obtained by refitting the model under the constraints that entries with index in $E'$ are 0.
\begin{definition}[Naive estimator]
	Given the loss function $\ell_n(X;\Theta)$ and the selected model $\mathcal{R}_E$, define the naive estimator as
	%\begin{equation*}
$
		\bar\Theta_E = \argmin_{\Theta\in \mathcal{R}_E} \left\{\ell_{n}(\Theta;X)\right\}.
$
%	\end{equation*}
\end{definition}

Let $\bar\theta_E=\vh(\bar\Theta_E)$. We refer to this estimator as \textit{naive} because, if the selected edge set $E$ were fixed in advance, its distribution would be asymptotically normal by the central limit theorem. However, since $E$ is data-dependent, inference based on this unconditional distribution is severely anti-conservative. This is established in the literature \citep{berk2013valid, fithian2017optimal} and confirmed by the simulations in Section~\ref{chapter3:sec:simulations}.

Another crucial quantity used in the construction of the randomization map is the estimator $\bar\theta^\perp_{E'}$.
\tcr{This estimator is asymptotically normal and orthogonal to $\bar\theta_E$ \citep[][Lemma 1]{guglielmini2025asymptotic}, and conditioning on its value helps eliminate nuisance parameters from the likelihood.}
%This follows from Lemma~1 in \citet{guglielmini2025asymptotic}, which proves this for a one-step estimator, which in turn is asymptotically equivalent to our naive estimator by the properties of one-step estimators \citep{lecam1956asymptotic}.}
Therefore, the two parameters are asymptotically independent.
\begin{definition}[Nuisance parameter]
	Given $A_E = H_{n,E'E}-J_{n,E'E} J^{-1}_{n,EE} H_{n,EE}$, define the nuisance parameter $\bar\theta^\perp_{E'}$ and its population version 
$\theta^{\perp*}_{E'}$ by
	\begin{equation*}
		\bar\theta^\perp_{E'} = n^{-1}\gradEc{\bar\Theta_E}-A_E\bar\theta_E 
        \mbox{ and } 
        \theta^{\perp*}_{E'} = n^{-1}\mathbb{E}\left[\gradEc{\Theta^*_E}\right]- A_{E} \theta^*_E.
	\end{equation*}
\end{definition}

Finally, in order to simplify our notation, we define the derivative of the lasso penalty term with respect to $\vh(\Theta)$. Define $\mathcal P'_{mat}(S,U)$ as the $p\times p$ symmetric matrix such that $\vh(\mathcal P'_{mat}(S,U)) = \mathcal P'_{vech}(S,U)$, where
\begin{equation*}
	\mathcal P'_{vech}(S,U)_j =
	\begin{cases}
		\sqrt{n}\lambda S_{k} & \text{if } j \in E_k, \\
		\sqrt{n}\lambda U_k & \text{if } j \in E'_k.
	\end{cases}
\end{equation*}
Then the penalty derivative is $\mathcal P'(S, U) = \frac{1}{2} D_p^\tp \, \vc(\mathcal P'_{mat}(S,U))$.
Denote
\begin{align*}
	\bar\theta^\perp = (0^\tp_{|E|}, \bar{\theta}^{\perp\tp}_{E'})^\tp,\; \theta^{\perp*} = (0^\tp_{|E|}, \theta^{\perp*\tp}_{E'})^\tp,\; C_1=-J_{n,E}J_{n,EE}^{-1}H_{n,EE},\; C_2 = H_{n,E}\diag(S).
\end{align*}

\begin{proposition}[Randomization map]\label{chapter3:map}
	Given the quantities $C_1$, $C_2$ and the function $\mathcal P'(s, u)$, for fixed $\theta$, $\theta^\perp$, define the map
	\begin{equation*}
		\Pi_{\theta, \theta^\perp}:\mathbb{R}^{|E|}\times\mathbb{R}^{d}\to \mathbb{R}^d:(v,u)\mapsto C_1\theta+C_2v +\mathcal P'(S, u) + \theta^\perp.
	\end{equation*}
	Then, the first-order conditions of the randomized graphical lasso problem in \eqref{chapter3:Theta_rho} can be represented as
%		\begin{equation*}
$
			\sqrt{n}\omega_n = \Pi_{\sqrt{n}\bar\theta_E,\sqrt{n}\bar\theta^\perp}(\sqrt{n}\widehat V,\widehat U) + o_P(1),
$
%	\end{equation*}
\end{proposition}
\tcr{Here and throughout, $o_P(1)$ denotes a term that converges to zero in probability as $n \to \infty$, where the asymptotic considerations are taken with respect to the true, unconditional distribution of the data.}
Note that the randomization map consists of quantities which are either fixed by conditioning or asymptotically equivalent to fixed population values.
Denote then the asymptotically equivalent vector
$\sqrt{n}\overline\omega_n = \Pi_{\sqrt{n}\bar\theta_E, \sqrt{n}\bar\theta^\perp}(\sqrt{n}\widehat V, \widehat U)$.
For randomized group lasso selection in the regression model, a similar representation was obtained by \citet{huang2025selective}  \citep[see also][]{panigrahi2023approximate_grouplasso}.
Here, the argument is developed for Gaussian graphical models with emphasis on the precision matrix, rather than a regression setting, resulting in a different parameterization and selection event.

\subsection{Selective likelihood}\label{chapter3:sec:inference}

%In this section, we derive the selective distribution of the parameters of interest, that is, their distribution conditional on the observed selection event. The resulting likelihood is valid for inference on the selected edges.

The starting point for the selective distribution is the distribution of the naive estimator $\bar\theta_E$, conditional on the selection event in Equation~\eqref{chapter3:event}, and additionally on the nuisance vector $\bar\theta^\perp_{E'}$:
\begin{equation}\label{chapter3:cond_distn}
	\sqrt{n}\bar\theta_E \Big\lvert \left\{\widehat S = S, \widehat U=U, \sqrt{n}\widehat V \in \mathbb{R}_+^{|E|}, \bar\theta^\perp_{E'} = \theta^\perp_{E'}\right\}.
\end{equation}
By conditioning on $\bar\theta^\perp_{E'}$ in this distribution, we eliminate the nuisance parameter $\theta^{\perp*}_{E'}$ from the likelihood, rendering it a function of $\theta^*_E$ alone.

To derive the conditional distribution, consider the conditional joint law of
\begin{equation*}
	\sqrt{n}(\bar\theta_E^\tp, \widehat V^\tp)^\tp  \Big\lvert \left\{\widehat S = S, \widehat U=U, \bar\theta^{\perp}_{E'} = \theta^\perp_{E'}\right\}.
\end{equation*}
Denote its likelihood by $L_{n,S,U,\theta^\perp_{E'}}(\theta^*_E, \theta^{\perp*}; \bar\theta_E, \widehat V)$. Truncating the likelihood to the event $\{\widehat V \in \mathbb{R}_+^{|E|}\}$, the conditional log-likelihood of interest is obtained:
% \begin{align*}
% 	\dfrac{L_{n,S,U,\theta^\perp_{E'}}(\theta^*_E, \theta^{\perp*}; \bar\theta_E, \widehat V)\mathbbm{1}_{\mathbb{R}_+^{|E|}}(\sqrt{n}\widehat V)}{\int L_{n,S,U,\theta^\perp_{E'}}(\theta^*_E, \theta^{\perp*}; \tilde{\theta}, \tilde{V})\mathbbm{1}_{\mathbb{R}_+^{|E|}}
% 		(\sqrt{n}\tilde{V})d\tilde{\theta}d\tilde{V}} =
% 	\dfrac{ L_{n,S,U,\theta^\perp_{E'}}(\theta^*_E, \theta^{\perp*}; \bar\theta_E, \widehat V)\mathbbm{1}_{\mathbb{R}_+^{|E|}}(\sqrt{n}\widehat V)}{\mathbb P
% 		[\sqrt{n}\widehat V\in\mathbb{R}_+^{|E|}\mid\bar\theta^\perp = \theta^\perp, \widehat S = S, \widehat U=U]},
% \end{align*}
% with log-likelihood
\begin{equation*}
	\log L_{n,S,U,\theta^\perp_{E'}}(\theta^*_E, \theta^{\perp*}; \bar\theta_E, \widehat V) - \log\mathbb P[\sqrt{n}\widehat V\in\mathbb{R}_+^{|E|}\mid\bar\theta^\perp = \theta^\perp, \widehat S = S, \widehat U=U].
\end{equation*}

We first identify the unconditional density in Lemma~\ref{chapter3:prop:density}.
\begin{lemma}[Unconditional density]\label{chapter3:prop:density}
	\tcr{Assume 
(i) $X_1,\ldots,X_n$ i.i.d.~$N_p(0_p,\Theta^{-1})$, where $p$ is fixed (ii) $\omega\sim N_d(0_d,\Omega)$,	
%	(i) the observations are i.i.d. from the model described in Section~\ref{chapter3:sec:randomized};\\
%	(ii) the randomization variable is drawn from a centered Gaussian distribution with known covariance matrix $\Omega$ as in Section~\ref{chapter3:sec:randomized};\\
	(iii) there is a sample size $n_0$ such that, for all $n\geq n_0$, the distribution of
%	\begin{equation*}
$
		\sqrt{n}((\bar\theta_E-\theta^*_E)^\tp, (\bar\theta^\perp_{E'}-\theta^{\perp*}_{E'})^\tp, \overline\omega_n^\tp)^\tp
$
%	\end{equation*}
	admits a Lebesgue density.}
	Define $\Sigma_{n,E} = H_{n,EE}^{-1}J_{n,EE}H_{n,EE}^{-1} \stackrel{P}{\to}\Sigma_E $ and $\Sigma_{n,E'}^\perp = J_{n,E'E'} - J_{n,E'E}J_{n,EE}^{-1} J_{n,EE'} \stackrel{P}{\to} \Sigma_{E'}^\perp$ as $n\to\infty$,
	% population values $\Sigma_E  = \lim_{n\to\infty}\Sigma_{n,E}$ and  $\Sigma_{E'}^\perp  = \lim_{n\to\infty}\Sigma_{n,E'}^\perp$, 
	\tcr{where the convergence is in probability}. 
Evaluated at the realized value  $(\sqrt{n}\theta_E^\tp, \sqrt{n}(\theta^{\perp}_{E'})^\tp, \sqrt{n}V^\tp, U^\tp)^\tp$,	
the asymptotic density of $(\sqrt{n}\bar\theta_E^\tp, \sqrt{n}(\bar\theta^{\perp}_{E'})^\tp, \sqrt{n}\widehat V^\tp, \widehat U^\tp)^\tp$  is proportional to
	\begin{equation*}
		\rho(\sqrt{n}\theta_E;\sqrt{n}\theta^*_E, \Sigma_E)\rho(\sqrt{n}\theta^\perp_{E'};\sqrt{n}\theta^{\perp*}_{E'}, \Sigma_{E'}^\perp)\rho(\Pi_{\sqrt{n}\theta, \sqrt{n}\theta^\perp}(\sqrt{n}V,  U); 0_d,  \Omega),
	\end{equation*}
	where $\rho(x;\mu,\Sigma)$ is the density of a multivariate normal variable with mean $\mu$ and covariance matrix $\Sigma$, evaluated at $x$.
\end{lemma}
\tcr{Note that, after conditioning on the selection event, the population values $\theta^*_E$ and $\theta^{\perp*}_{E'}$ are fixed.}

Truncating the unconditional density to the selection event yields the selective likelihood, which is derived next.
\begin{theorem}[Selective likelihood]\label{chapter3:theorem:conddensity}
	\tcr{Assume the conditions of Lemma~\ref{chapter3:prop:density} hold.}
	The law of
	\begin{equation}\label{chapter3:eq:density}
		\sqrt{n}(\bar\theta^\tp_E, \widehat V^\tp)^\tp | \{\widehat S = S, \widehat U = U, \bar\theta^\perp = \theta^\perp\},
	\end{equation}
	truncated to the event $\{\widehat V\in\mathbb{R}_+^{|E|}\}$ is proportional to
	\begin{equation}\label{chapter3:eq:condlikelihood}
		\dfrac{\rho(\sqrt{n}\bar\theta_E; L\sqrt{n}\theta^*_E+m, Z)\rho(\sqrt{n}\widehat V; P\sqrt{n}\bar\theta_E+q,\Delta)\mathbbm{1}_{\mathbb{R}_+^{|E|}}(\sqrt{n}\widehat V)}{\int \rho(\sqrt{n}\tilde{\theta}; L\sqrt{n}\theta^*_E+m, Z)\rho(\sqrt{n}\tilde{V}; P\sqrt{n}\tilde{\theta}+q,\Delta)\mathbbm{1}_{\mathbb{R}_+^{|E|}}(\sqrt{n}\tilde{V})d\tilde{\theta}d\tilde{V}}
	\end{equation}
	where
%	\begin{align*}
$
		\Delta = (C_2^\tp\Omega^{-1}C_2)^{-1}; \quad P = -\Delta C_2^\tp\Omega^{-1}C_1; \quad q = -\Delta C_2^\tp\Omega^{-1}(\mathcal P'(S,U)+\sqrt{n}\bar\theta^\perp); %\\
		Z = (\Sigma_E^{-1} - P^\tp\Delta^{-1}P + C_1^\tp\Omega^{-1}C_1)^{-1}; \quad L = Z\Sigma_E^{-1}; %\\ 
		m = Z(P^\tp\Delta^{-1}q-C_1^\tp\Omega^{-1}(\mathcal P'(S,U)+\sqrt{n}\bar\theta^\perp)).
$
%	\end{align*}
\end{theorem}

\subsubsection{Computation of the normalizing constant}
%In Theorem~\ref{chapter3:theorem:conddensity}, we derive an expression for the selective likelihood. 
Computing the selective likelihood requires calculating the normalizing constant, which is not available in closed form.
Following \citet{huang2025selective}, we approximate the integral by its Laplace approximation
\begin{align*}
	& \inf_{\tilde\theta,\tilde V \in \mathbb{R}_+^{|E|}}\Big\{\frac{1}{2}(\sqrt{n}\tilde\theta-L\sqrt{n}\theta^*_E-m)^\tp Z^{-1}(\sqrt{n}\tilde\theta-L\sqrt{n}\theta^*_E-m) \\
	&\;\;\;\;\;\;\;\;\;\;\;\;+ \frac{1}{2}(\sqrt{n}\tilde V-P\sqrt{n}\theta-q)^\tp\Delta^{-1}(\sqrt{n}\tilde V-P\sqrt{n}\theta-q)\Big\}.
\end{align*}
We solve an unconstrained version of the problem by introducing a barrier function $\phi(\nu)=-\sum_{j=1}^{|\nu|}\log(\nu_j)$ in the optimization objective, which enforces the linear constraints $\nu_j > 0$.
% \begin{align*}
% 	& \inf_{\tilde\theta,\tilde V}\Big\{\frac{1}{2}(\sqrt{n}\tilde\theta-L\sqrt{n}\theta^*_E-m)^\tp Z^{-1}(\sqrt{n}\tilde\theta-L\sqrt{n}\theta^*_E-m) \\
% 	&\;\;\;\;\;\;\;\;\;+ \frac{1}{2}(\sqrt{n}\tilde V-P\sqrt{n}\theta-q)^\tp\Delta^{-1}(\sqrt{n}\tilde V-P\sqrt{n}\theta-q)+\phi(\sqrt{n}\tilde V)\Big\}.
% \end{align*}
The resulting log-likelihood then equals
\begin{align}\label{chapter3:loglik}
	\log\rho(&\sqrt{n}\bar\theta_E; L\sqrt{n}\theta^*_E+m, Z)+\frac{1}{2}\inf_{\tilde\theta,\tilde V}\Big\{(\sqrt{n}(\tilde\theta-L\theta^*_E)-m)^\tp Z^{-1}(\sqrt{n}(\tilde\theta-L\theta^*_E)-m) \nonumber \\
	&+ (\sqrt{n}\tilde V-P\sqrt{n}\theta-q)^\tp\Delta^{-1}(\sqrt{n}\tilde V-P\sqrt{n}\theta-q)+\phi(\sqrt{n}\tilde V)\Big\}.
\end{align}

\subsection{Selective maximum likelihood inference}

We now define the selective MLE $\widehat\theta^{SMLE}_E$ and the Fisher information matrix. % $[\widehat\Sigma_E^{SMLE}]^{-1}$.

\begin{theorem}[Selective MLE and Fisher information]\label{chapter3:theorem:mle}
	Using the definitions \tcr{and under the same assumptions} as in Theorem~\ref{chapter3:theorem:conddensity}, the selective MLE is
	\begin{equation*}
		\sqrt{n}\widehat\theta^{SMLE}_E = L^{-1}\sqrt{n}\bar\theta_E + L^{-1}Z P^\tp\Delta^{-1}(P\sqrt{n}\bar\theta_E + q - \sqrt{n} \widehat V^{SMLE}) - L^{-1}m,
	\end{equation*}
	where $\widehat V^{SMLE} = \argmin_{\tilde{V}}\{\frac{1}{2}(\sqrt{n}\tilde{V}-P\sqrt{n}\bar\theta_E-q)^\tp\Delta^{-1}(\sqrt{n}\tilde{V}-P\sqrt{n}\bar\theta_E-q) + \phi(\sqrt{n}\tilde{V})\}$, and the observed Fisher information $[\widehat\Sigma_E^{SMLE}]^{-1}=$ 
	\begin{equation*}
%		[\widehat\Sigma_E^{SMLE}]^{-1} = 
		L^{-1}Z L^{-\tp} + L^{-1}Z [P^\tp\Delta^{-1}P - P^\tp\Delta^{-1}(\Delta^{-1}+\nabla^2\phi(\widehat V^{SMLE}))^{-1}\Delta^{-1}P]Z L^{-\tp}.
	\end{equation*}
\end{theorem}

%Using the SMLE and observed Fisher information from 
Theorem~\ref{chapter3:theorem:mle} provides a normal approximation for inference on $\theta^*_E$: 
\begin{empheq}[box=\graybox]{equation}
\label{chapter3:original}
	\sqrt{n}\,(\widehat\theta^{SMLE}_E - \theta^*_E) \dot\sim N_{|E|}\bigl(0_{|E|},\widehat\Sigma_E^{SMLE}\bigr).
\end{empheq}
The distribution in \eqref{chapter3:original} can directly be used to construct post-selection confidence intervals and hypothesis tests, as it contains all selection information.

\noindent\textbf{Differentiable functions}.
Let $\widehat\Theta^{SMLE}_E$ denote the matrix form of the SMLE, defined so that
$\vh(\widehat\Theta^{SMLE}_E)_E = \widehat\theta^{SMLE}_E$ and $\vh(\widehat\Theta^{SMLE}_E)_{E'} = 0_{|E'|}$.

Given a differentiable function $T_{\text{diff}}(\cdot)$ with estimator $T_{\text{diff}}(\widehat\theta^{SMLE}_E)$ on the edges in the selected graph and target $T_{\text{diff}}(\theta^*_E)$, by the multivariate delta method applied to the distribution of the SMLE (\textit{selective delta method}),
\begin{empheq}[box=\graybox]{equation*}
	\sqrt{n}\left(T_{\text{diff}}(\widehat\theta^{SMLE}_E)-T_{\text{diff}}(\theta^*_E)\right) \dot\sim N\left(0,T'_{\text{diff}}(\widehat\theta^{SMLE}_E)^\top \widehat\Sigma_E^{SMLE} T'_{\text{diff}}(\widehat\theta^{SMLE}_E)\right),
\end{empheq}
where $T'_{\text{diff}}(\theta)$ is the derivative with respect to $\theta$.

As an example %of a differentiable function, 
we consider the partial correlations between nodes which are connected in the target graph.
Given %a precision matrix 
$\Theta$, the partial correlation between nodes $i$ and $j$ is $\rho_{ij} = -\frac{\Theta_{ij}}{\Theta_{ii}\Theta_{jj}}$. We take $T_{\text{diff}}(\cdot)$ that maps $\theta$ into $\rho_{ij}$.
Note that $T'_{\text{diff}}$ consists of the vector-half of the matrix with entries $\frac{\partial\rho_{ij}}{\partial\Theta_{kl}}$, where
\begin{equation*}
	\frac{\partial\rho_{ij}}{\partial\Theta_{kl}} =
	\begin{cases}
		- \frac{1}{\sqrt{\Theta_{ii}\Theta_{jj}}} \quad\text{if } (k,l)=(i,j); \\
		 \frac{\Theta_{ij}}{2\sqrt{\Theta_{ii}^3\Theta_{jj}}} \quad\text{if } (k,l)=(i,i); \\
			 \frac{\Theta_{ij}}{2\sqrt{\Theta_{ii}\Theta_{jj}^3}} \quad\text{if } (k,l)=(j,j); \\
			0 \quad\text{otherwise.}
	\end{cases}
\end{equation*}

\noindent\textbf{Non-differentiable functions}.
Consider a function $T(\cdot)$, not necessarily differentiable, with corresponding estimator $T(\widehat\theta^{SMLE}_E)$ which is $\sqrt{n}$-bounded.

We aim to test the hypothesis $H_0 : T(\theta^*_E) = T_0$.
Let $F_0(t)$ denote the cumulative distribution function of $T$ under $H_0$:
\begin{equation}\label{chapter3:T_cdf}
	F_0(t) = \int_{\{\theta : T(\theta) \leq t\}} \rho\left(\sqrt{n} \theta,\, \sqrt{n} \theta_0,\, \widehat\Sigma_E^{SMLE}\right) d\theta,
\end{equation}
where $\theta_0 \in \{\theta : T(\theta) = T_0\}$, i.e. $\theta_0$ is a value such that the null hypothesis is satisfied.
In practice, we set $\theta_0 = \widehat\theta_0$, where $\widehat\theta_0$ is a parameter value that satisfies the null while remaining as close as possible to the observed SMLE. In other words, $\widehat\theta_0$ is the projection of $\widehat\theta^{SMLE}_E$ onto the parameter space constrained by $H_0$, minimizing the Mahalanobis distance defined by the selective covariance.
We formalize this in Equation~\eqref{chapter3:theta_0}.

By construction, $F_0(T(\widehat\theta^{SMLE}_E))$ has a null distribution that is uniform on $(0,1)$ and is thus a pivotal quantity for $H_0$.
Computing the integral in \eqref{chapter3:T_cdf} is infeasible in practice. In order to estimate it, we use a restricted parametric bootstrap (\textit{selective bootstrap}) approach:
\begin{grayenum}
\begin{enumerate}
	\item Obtain $\widehat\theta_0$ as the solution to the constrained optimization problem
	\begin{equation}\label{chapter3:theta_0}
		\sqrt{n} \hat\theta_0 = \argmin_{\theta: T(\theta) = T_0}
		\left\{ (\widehat\theta^{SMLE}_E - \theta)^\top [\widehat\Sigma_E^{SMLE}]^{-1} (\widehat\theta^{SMLE}_E - \theta) \right\},
	\end{equation}
	which is solved numerically using the COBYLA algorithm via the R package \texttt{nloptr} \citep{johnson2007nlopt,powell1994cobyla}.
	\item Generate $N$ samples $\sqrt{n}\theta_0^{(h)} \sim N(\sqrt{n}\hat\theta_0,\, \widehat\Sigma_E^{SMLE})$, $h = 1,\dots,N$, and compute $T_0^{(h)} = T(\theta_0^{(h)})$ for each of them.
	\item Compute the p-value:
	
	- If $T_0 = 0$ (one-sided test):
	$p = \frac{\#\{T_0^{(h)} \le T(\widehat\theta^{SMLE}_E)\}}{N}.$
	
	- If $T_0 \neq 0$ (two-sided test): \\
	\hspace*{0.3cm}$p = 2 \min\left\{
	\frac{\#\{T_0^{(h)} \le T(\widehat\theta^{SMLE}_E)\}}{N},
	\frac{\#\{T_0^{(h)} \ge T(\widehat\theta^{SMLE}_E)\}}{N}
	\right\}.$
\end{enumerate}
\end{grayenum}

This procedure yields a valid post-selection test for non-differentiable functionals. Node strength is a relevant example of such functions, since, by Lemma~\ref{chapter3:strength_boundedness} it satisfies the assumption of $\sqrt{n}$-boundedness.
The same technique holds to obtain a p-value for bridge strength or for the difference in node (or bridge) strength between two nodes. These measures may be computed using the precision matrix entries or the partial correlations as weights.

\begin{lemma}[$\sqrt{n}$-boundedness of the node strength estimator]\label{chapter3:strength_boundedness}
	Consider $\sqrt{n}X\sim N(\mu, C)$, where $\mu$ and $C$ are $O(1)$, an arbitrary set $E$, and the node strength estimator $T(X) = \sum_{i\in E}|X_i|$. Then the estimator $T(X)$ is $\sqrt{n}$-bounded.
\end{lemma}

For simplicity, we present the method using precision matrix elements as edge weights. Replacing $\widehat{\theta}^{SMLE}_E$ with the estimated partial correlations and $\widehat{\Sigma}^{SMLE}_E$ with their delta method covariance matrix allows the use of the more informative partial correlations as edge weights. These are employed in the simulations and data application in Sections~\ref{chapter3:sec:simulations} and \ref{chapter3:sec:application}.

\subsection{Elastic net penalty}\label{chapter3:sec:elasticnet}

When nodes are highly correlated with each other, estimation can become unstable \citep{kovacs2021graphical,zou2005regularization}. Similarly to elastic net regression, an additional ridge penalty can be applied in the graphical lasso problem to mitigate this instability. \tcr{Furthermore, the ridge term ensures the existence of a solution also in the randomized setting (Proposition~\ref{chapter3:prop:existence_elnet}). Note that a very small ridge penalty is sufficient to guarantee the existence of a solution, and the penalty can be chosen to be asymptotically negligible.}

%The penalty is thus $\lambda [\gamma x + \frac{1-\gamma}{2}x^2]$, where $\lambda>0$, $\lambda=O_P(n^{-1/2})$, $\gamma \in [0,1]$ and $\gamma=O_P(1)$.

The ``graphical elastic net'', defined in Equation~\ref{chapter3:elnet_objective} can be solved with the algorithm in \citet{kovacs2021graphical}, which uses coordinate descent in a modified graphical lasso algorithm.
\tcr{ Using the Frobenius norm $\|\Theta\|_F^2 = \sum_{i,j}\Theta_{ij}^2$,
\begin{equation}\label{chapter3:elnet_objective}
	\widehat\Theta^{\text{EN}} = \argmin_{\Theta \succ 0} \left\{\tr((C_n-W_n)\Theta)-\log\det(\Theta) +\lambda\gamma||\Theta||_1 + \frac{\lambda(1-\gamma)}{2}\|\Theta\|_F^2\right\}.
\end{equation}
\begin{proposition}\label{chapter3:prop:existence_elnet}
	For any $\lambda>0$ and $\gamma \in (0,1)$, the solution to the graphical elastic net problem in Equation~\eqref{chapter3:elnet_objective} exists and is unique.
\end{proposition}
}

The method for selective inference after elastic net selection 
%is analogous to the one for the lasso selection, with 
requires 
an adjustment to the definition of the randomization map.
Define $\tilde C_0 = \frac{1}{2}\lambda(1-\gamma)D_p^\tp D_p$ and $\tilde C_2 = (H_{n,E}+\tilde C_{0,E})\diag(S)$.

\begin{proposition}\label{chapter3:map_elasticnet}
	Given the quantity $\tilde C_2$ define above and $C_1$, $\tilde{\mathcal P}'(s, u)$ as in Proposition~\ref{chapter3:map}, for fixed $\theta$, $\theta^\perp$, define the map
	\begin{equation*}
		\tilde\Pi_{\theta,\theta^\perp}:\mathbb{R}^{|E|}\times\mathbb{R}^{d}\to \mathbb{R}^d:(v,u)\mapsto C_1\theta+\tilde C_2b +\gamma\mathcal P'(S, u) + \theta^\perp.
	\end{equation*}
	The first-order conditions of the randomized graphical elastic net yield
	\begin{equation*}
		\sqrt{n}\omega_n = \tilde\Pi_{\sqrt{n}\bar\theta_E,\sqrt{n}\bar\theta^\perp}(\sqrt{n}\widehat V,\widehat U) + o_P(1).
	\end{equation*}
\end{proposition}

When working with the elastic net penalty, the definitions in Proposition~\ref{chapter3:map} should be substituted by the ones in Proposition~\ref{chapter3:map_elasticnet}. Note that the adjustments to the matrices only depend on fixed quantities. Proposition~\ref{chapter3:prop:density}, Theorems~\ref{chapter3:theorem:conddensity} and \ref{chapter3:theorem:mle} still hold, with the definitions in Proposition~\ref{chapter3:map_elasticnet}.

\section{Simulation study}\label{chapter3:sec:simulations}

We conduct a simulation study to evaluate the performance of our method for selective inference after graphical model selection.

For the randomized approach, we use $\Omega = \sigma^2_\omega I_d$, we set $\lambda=\sqrt{2\sigma_\omega\log(p)/n}$ and for the non-randomized approach, we set $\lambda=\sqrt{\log(p)/n_1}$, where $n_1$ is the size of the sample used for selection in data splitting. We take $\lambda=\sqrt{\log(p)/n}$ for the polyhedral and naive methods. These choices follow from empirical considerations, with the goal of obtaining a similar number of selected edges across methods for a fair comparison in this simulation study.

Simulations show that these values lead to similar levels of selection, which allows a fair comparison of the selection accuracy across methods.
We generate data from a centered and scaled multivariate normal distribution and conduct simulations using both the graphical lasso penalty and the elastic net with $\gamma=0.5$.
We compare the following methods in each run:
\begin{enumerate}
	\item \textit{Selective maximum likelihood (SMLE):} our method for selective inference with randomization, with $\sigma_\omega$ varying between 0.5 and 1.25;
	\item \textit{Data splitting (DS):} using 10\%, 50\% or 90\% of the sample for selection and the rest for inference with the unconditional Gaussian pivot;
	\item \textit{Polyhedral (POLY):} selective inference with the polyhedral lemma (without randomization) \citep{guglielmini2025asymptotic};
	\item \textit{Naive:} unconditional Gaussian pivot, ignoring selection.
\end{enumerate}

The data are generated from a graph with $p=70$ nodes using the function \texttt{sample\_islands} from the R package \texttt{igraph} and adding random weights between 0.5 and 2.0, with random sign. The graph has a modular structure, i.e. it can be divided into 5 subgraphs which have higher density of connections within than between them. In particular, each community forms a subgraph with edges drawn with probability 0.3; between any two communities, 3 edges connect them. To ensure positive definiteness, we update diagonal entries as $\Theta_{ii} = \Theta_{ii} + \sum_{j\neq i}|\Theta_{ij}|$ for $i=1,\dots,p$. 
We compute inference for the selected edges, for the bridge strength (BS) and for the difference in bridge strength (BSD) between two randomly selected nodes. 

\subsection{Evaluation of performance}

We %evaluate the performance of the selection method, to 
measure how the accuracy of selection is affected by randomization and sample splitting. Since non-selected edges are not considered for inference, we focus on controlling the rate of discarded true positives. We report Recall as a measure of selection accuracy, defined as the proportion of selected true edges: $\text{Recall} = TP/(TP+FN)$, where $TP$ is the number of true positives and $FN$ the number of false negatives. 
To evaluate the quality of inference, we compute confidence intervals for all selected precision matrix elements (indicated as $\theta$) and measure the simulated coverage as the percentage of times that the target $\theta^*_j$, for $j \in E$ is included in the interval, comparing this with the nominal coverage.
Using the tests described in Section~\ref{chapter3:sec:inference}, we show the coverage for partial correlations and for partial correlation-based node strength (NS) for each node which is connected to at least one selected edge, and for the difference in node strength (NSD) between two randomly selected, non-isolated nodes. We also display average confidence interval length for the precision matrix elements and partial correlations.

\subsection{Results}

\paragraph{Selection accuracy and inferential power trade-off}

\begin{figure}[ht]
	\centering
	\includegraphics[width=0.45\linewidth]{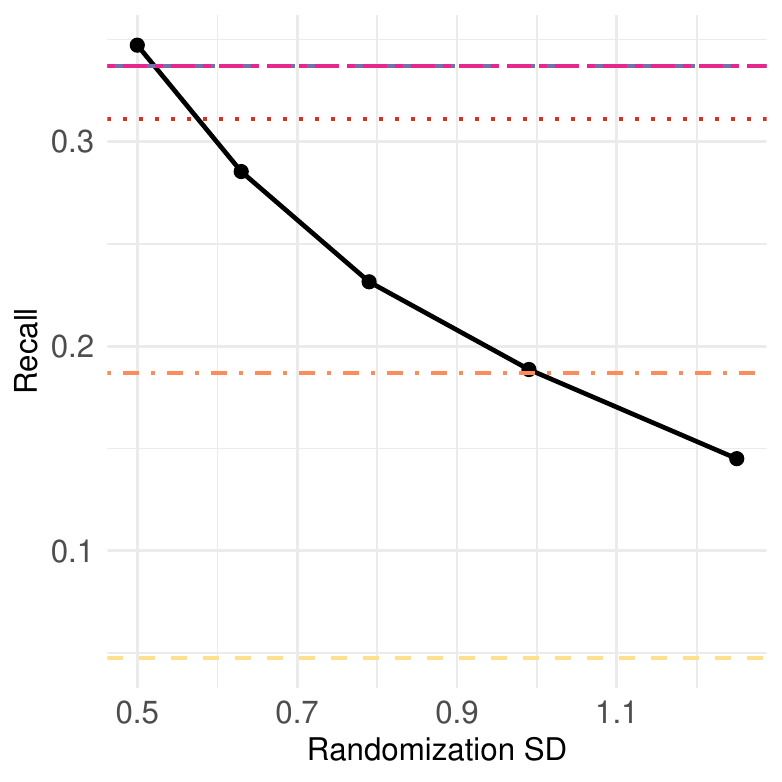}
	\includegraphics[width=0.45\linewidth]{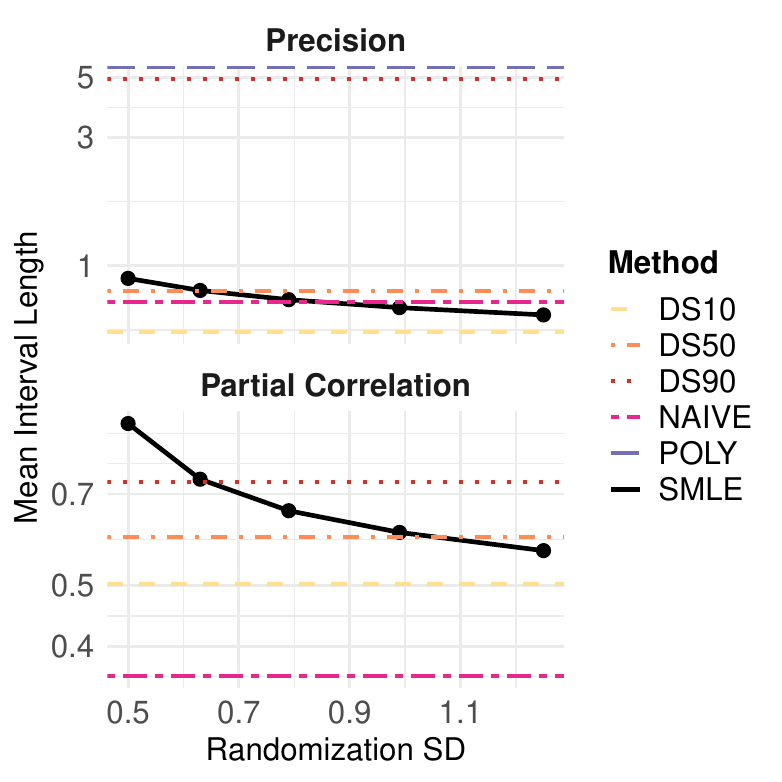}
	\caption{Left: Recall of the selection step for randomized graphical lasso, as the amount of randomization varies, compared to the other methods. Right: Average length of the confidence intervals for the precision and partial correlation elements for randomized graphical lasso as the amount of randomization varies, compared to the other methods. Each point is the average over 500 runs, with $n = 60$ and $p=70$.}
	\label{chapter3:fig:recall_sd}
\end{figure}

The left panel in Figure~\ref{chapter3:fig:recall_sd} displays the recall of the selection step for randomized graphical lasso and elastic net when $\sigma_\omega$ varies between 0.5 and 1.25, compared to the non-randomized selection method and data splitting with different proportions of the sample used for selection. At a low amount of randomization, our method achieves recall comparable to the full data selection. In terms of selection, this is a similar performance to data splitting with 90\% of the sample used for selection, which however does not reach nominal coverage for inference.
Increasing the amount of randomization seems empirically to lead to more edges being selected: this improves recall and avoids discarding true edges, but may also lead to selecting more false edges. Further understanding of how the amount of randomization affects the trade-off between true and false positives remains an open question.

The right panel shows the average length of the confidence intervals for the precision matrix elements and partial correlations for the different methods. The polyhedral method presents infinite average length for both targets, confirming the aforementioned theoretical results. As $\sigma_\omega$ increases, the SMLE yields shorter confidence intervals and thus more powerful inference. For the precision matrix elements, the average length is similar or lower than that of data splitting with a 50\% splitting proportion, while achieving better selection. For partial correlations, depending on the amount of randomization, the average length is comparable to either that of 50\% or 90\% data splitting, while attaining nominal coverage.

Analogous results for elastic net penalty are presented in the Appendix.

\paragraph{Asymptotic behavior}

\begin{figure}[ht]
	\centering
	\includegraphics[width=\linewidth]{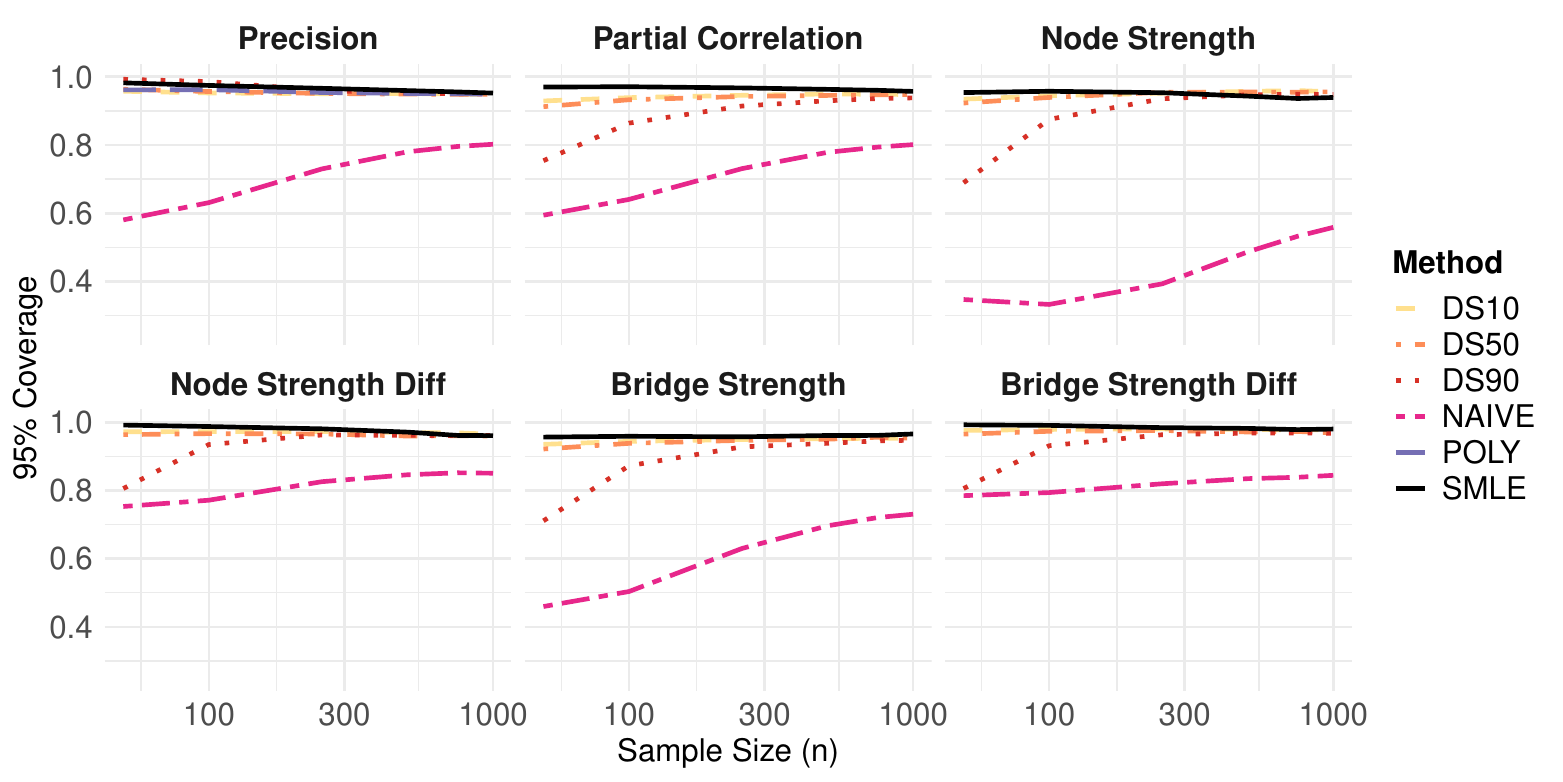}
	\caption{Coverage results for randomized graphical lasso with randomization standard deviation equal to 0.5 as the sample size $n$ varies, compared to the other methods. Each point is the simulated coverage over 500 runs, with $p=70$, for different targets. The nominal coverage is 95\%.}
	\label{chapter3:fig:coverage_n}
\end{figure}

We then investigate the performance of our method with randomization standard deviation equal to 0.5 as the sample size increases.
Figure~\ref{chapter3:fig:coverage_n} shows the coverage results for the different methods as the sample size $n$ varies, with $p=70$, for the graphical lasso penalty. The naive method presents under-coverage for all sample sizes, while data splitting with 90\% of the sample used for selection shows under-coverage at low sample sizes. The SMLE with randomization attains the nominal coverage across all targets and sample sizes, while data splitting with 10\% or 50\% of the sample used for selection also achieves satisfactory coverage. Note, however, that the coverage is defined for the selected target. Therefore, coverage for the selected target translates to better coverage to the actual true full graph if the selection is more accurate.

\begin{figure}[ht]
	\centering
	\includegraphics[width=0.45\linewidth]{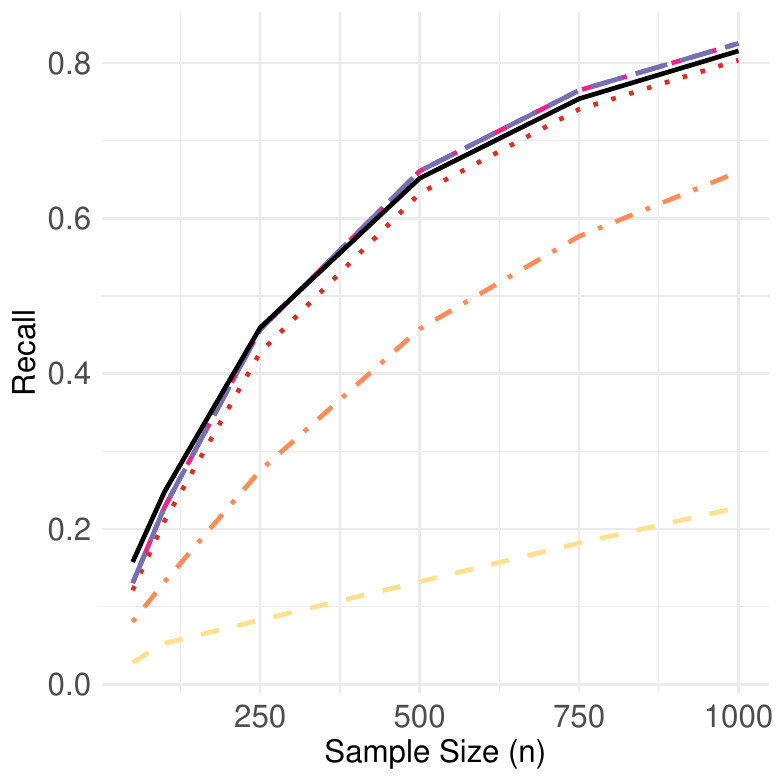}
	\includegraphics[width=0.45\linewidth]{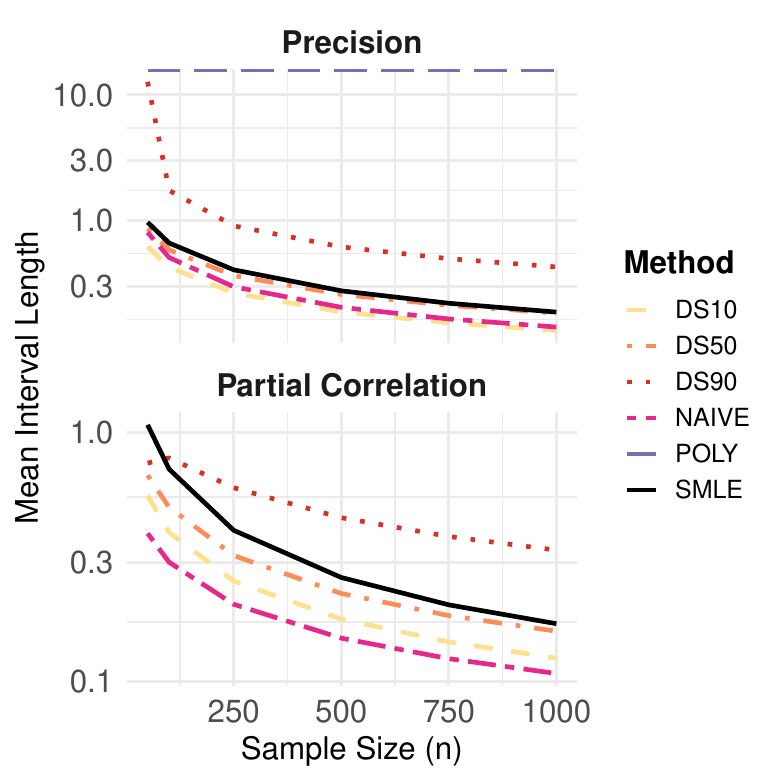}
	\caption{Left: Recall of the selection step for randomized graphical lasso with randomization standard deviation equal to 0.5 as the sample size $n$ varies, compared to the other methods. Right: Average length of the confidence intervals for the precision and partial correlation elements for randomized graphical lasso with randomization standard deviation equal to 0.5, as the sample size $n$ varies, compared to the other methods. Each point is the average over 500 runs, with $p=70$.}
	\label{chapter3:fig:recall_n}
\end{figure}

Figure~\ref{chapter3:fig:recall_n} shows that the recall of the selection step for randomized graphical lasso improves as the sample size increases, achieving similar or better performance than data splitting with 90\% of the sample used for selection.

\section{Application to brain imaging data}\label{chapter3:sec:application}

We use cortical thickness data, measured via magnetic resonance imaging (MRI), from $n=59$ healthy subjects in the OASIS-1 dataset \citep{marcus2007open}. The data, preprocessed using the software FreeSurfer, were provided by OASIS (Open Access Series of Imaging Studies). The resulting dataset consists of $p=66$ brain regions, parcellated according to the Desikan-Killiany atlas \citep{desikan2006automated} (excluding the \textit{corpus callosum}).
We model the data with a structural connectivity network and select the edges using the randomized graphical lasso, with $\lambda=\sqrt{\log(p)/n}$ and $\sigma_\omega=0.5$. %standard deviation of the randomization variable equal to 0.5. 

To illustrate the inferential targets described in Section~\ref{chapter3:sec:targets}, we consider hypotheses motivated by previous findings in the literature. In particular, the left precentral gyrus has been identified as one of the most central regions, while the
parahippocampal gyrus has weaker connectivity \citep{chen2008revealing}. Furthermore, the precuneus is strongly connected to multiple other regions \citep{margulies2009precuneus}, while the pericalcarine cortex typically exhibits weaker global connectivity \citep{sepulcre2010organization}. Although these results concern functional connectivity, structural and functional connectivity are often correlated \citep{greicius2009resting}.
For the analysis of bridge centrality, we consider 5 modules, according to the anatomical structure of the cortex: the frontal, parietal, temporal, occipital and cingulate lobes.

\begin{table}[H]
\label{chapter3:tab:dataapplication}
\caption{Selective inference on OASIS-1 cortical thickness data (n = 59, p = 66)}
\begin{tabular}{lrrrrr}
\textbf{Contrast} & \textbf{Region 1} & \textbf{Region 2} & \textbf{Est. 1} & \textbf{Est. 2} & \textbf{p-value} \\
\midrule
NS diff. & L. precentral & R. parahippocampal & 20.233 & 3.319 & $0$ \\
NS diff. & L. precuneus & L. pericalcarine & 9.232 & 1.142 & 0.006 \\
BS diff. & L. precentral & R. parahippocampal & 7.477 & 0.960 & 0.0002 \\
BS diff. & L. precuneus & L. pericalcarine & 4.350 & 0.803 & 0.2012 \\
\end{tabular}
\end{table}

The results are summarized in Table~\ref{chapter3:tab:dataapplication}. The conclusions drawn using the proposed method are generally in line with the previous literature on brain connectivity and display the conservative quality that is expected when accounting for selection.

\nocite{huber1981robust}


\bibliographystyle{plainnat}
\bibliography{main}

\begin{appendices}
\section{Lemma's and proofs}

\begin{lemma}\label{chapter3:density_formulation}
	The density $\rho(\sqrt{n}\theta_E;\sqrt{n}\theta^*_E, \Sigma_E)\rho(\Pi(\sqrt{n}V, U); 0_d, \Omega)$
	can be rewritten as
%	\begin{equation*}
$
		{d}(\theta^*_E)\rho(\sqrt{n}\theta_E; L\sqrt{n}\theta^*_E+ m, Z)\rho(\sqrt{n}V; P\sqrt{n}\theta_E+ q, \Delta),
$
%	\end{equation*}
	where
	%  $c_E = 1/[(2\pi)^{|{E}|+d}|{\Sigma_E}|| \Omega|]^{1/2}$, $c(\theta^*_E) =  q^\tp \Delta^{-1} q-\frac{1}{2}(\mathcal P'(S,U)+\sqrt{n}\theta^\perp)^\tp \Omega^{-1}(\mathcal P'(S,U)+\sqrt{n}\theta^\perp) - \frac{1}{2}n\theta^{*\tp}_E{\Sigma_E}^{-1}\theta^*_E$ and $d(\theta^*_E) = c_E\exp\{-(L\sqrt{n}\theta^*_E+m)^\tp Z^{-1}( L\sqrt{n}\theta^*_E+ m)+c(\theta^*_E)\}$ 
	$c_E$, $c(\theta^*_E)$ and $d(\theta^*_E)$ are constants which do not depend on $\theta_E$ or $V$.
\end{lemma}
\begin{proof}[Proof of Lemma~\ref{chapter3:density_formulation}]
	The result follows from the algebraic decomposition in the proof of
Proposition~2.4 in \citet{huang2025selective}.
\end{proof}

\begin{proof}[Proof of Proposition~\ref{chapter3:rglasso_convexity}]
	The following results are established in Chapter 3 of \citet{boyd2004convex}. The function $-\log\det(\Theta)$ is strictly convex over the cone of symmetric positive definite matrices. The $\ell_1$ norm is convex, hence so is $\lambda\|\Theta\|_1$. Since the sum of a strictly convex function and convex functions is strictly convex, the objective function is strictly convex. 
\end{proof}

\begin{proof}[Proof of Lemma~\ref{chapter3:hj_posdef}]
	By \cite{guglielmini2025asymptotic}\\
	$H_n(\Theta) = n^{-1}\sum_{h=1}^n \frac{\partial^2 \ell(\Theta,X_h)}{\partial\theta\partial\theta^\tp} = \frac{1}{2}D_p^\tp(\Theta^{-1}\otimes\Theta^{-1})D_p$.
	Since $\Theta$ is positive definite, so is $\Theta^{-1}$. As the Kronecker product of positive definite matrices, $\Theta^{-1}\otimes\Theta^{-1}$ is thus positive definite.
	The duplication matrix has full column rank \citep[Equation (4.4) in][]{harville1998matrix}, therefore, by Theorem 14.2.9 in \citet{harville1998matrix}, $D_p^\tp(\Theta^{-1}\otimes\Theta^{-1})D_p$ is positive definite. Noting that multiplication by a scalar does not affect positive-definiteness, this proves that $H_n(\Theta)$ is positive definite.
	Any principal submatrix of a positive definite matrix is positive definite, so $H_{n,EE}(\Theta)$ is positive definite by Corollary 14.2.12 in \citet{harville1998matrix}.
	
	For $J_n(\Theta)$, obtain the $1 \times d$ derivative
			$\frac{\partial\ell(\Theta; X_h)}{\partial \theta^\tp} = \frac{1}{2}\vc(-\Theta^{-1}+X_hX_h^\tp)^\tp D_p$,
		and define $G_n$ as the $n\times d$ matrix in which the $h$-th row is as defined above. Then $J_n(\Theta) = n^{-1}G_n^\tp G_n$, and thus is positive definite whenever $\text{rank}(G_n)=d$, by Corollary 14.2.14 in \citet{harville1998matrix}. This is true if $n\geq d$, and in this case $J_n(\Theta)$ is positive definite and also its submatrix $J_{n,EE}(\Theta)$.
		
	Define the $n\times |E|$ matrix $G_{n,E}$ as the set of columns of $G_n$ indexed by $E$.
	Then $J_{n,EE}(\Theta) = n^{-1}G_{n,E}^\tp G_{n,E}$, and therefore, if $\text{rank}(G_{n,E}) = |E|$, the matrix $J_{n,EE}(\Theta)$ is positive definite by Corollary 14.2.14 in \citet{harville1998matrix}.
	\end{proof}

\begin{proof}[Proof of Proposition~\ref{chapter3:map}]
	By the K.K.T. conditions of the optimization problem in \eqref{chapter3:Theta_rho}
	\begin{align*}
		&\frac{\partial}{\partial\theta}\left\{\tr((C_n-W_n)\Theta)-\log\det(\Theta) +\lambda||\Theta||_1\right\}\mid_{\Theta=\widehat\Theta} = 0 \\
		% &D_p^\tp\vc(C_n-W_n-\widehat\Theta^{-1}+\frac{1}{\sqrt{n}}\mathcal P'_{mat}(S, U))=0 \\
		% &\frac{\sqrt{n}}{2}D_p^\tp\vc(C_n-\widehat\Theta^{-1})-\frac{\sqrt{n}}{2}D_p^\tp\vc(W_n)+\frac{1}{2}D_p^\tp \vc(\mathcal P'_{mat}(S, U))=0, \\
		&n^{-1/2}\grad{\widehat\Theta}-\sqrt{n}\omega_n+\mathcal P'(S, U)=0.
	\end{align*}
	Now take a Taylor series expansion of the gradient evaluated at the regularized estimator around the naive estimator:
	\begin{align*}
		n^{-1/2}\grad{\widehat\Theta}
		= n^{-1/2}\grad{\bar\Theta_E} + H_{n,E}\sqrt{n}(\widehat\theta_E-\bar\theta_E) + o_P(1).
	\end{align*}
	Then, using the definition of the orthogonal nuisance vector, and the fact that
	$\widehat\theta_E = \diag(S)V$, the K.K.T. conditions can be written as
$		\sqrt{n}\omega_n 
		= C_1\sqrt{n}\bar\theta_E + C_2\sqrt{n}V + \mathcal P'(S,U) + \sqrt{n}\bar\theta^\perp.
$
\end{proof}

\begin{proof}[Proof of Lemma~\ref{chapter3:prop:density}]
	Consider the transformation $\Pi^*:(\theta,\theta^\perp,\omega) \mapsto (\theta,\theta^\perp,\Pi^{-1}_{\theta,\theta^\perp}(\omega))$, so that $(V^\top,U^\top)^\top=\Pi^{-1}_{\theta,\theta^\perp}(\omega)$.
	Denote by $\bar g_n$ the density of $\sqrt{n}(\bar\theta_E^\tp, (\bar\theta_{E'}^\perp)^\tp,\overline\omega_n^\tp)^\tp$ and apply the transformation, obtaining the density of $(\sqrt{n}\theta_E,\sqrt{n}\theta^\perp_{E'},\sqrt{n}V,U)$ which is proportional to
	\begin{align*}
		|\det D_{(\Pi^*)^{-1}}&(\sqrt{n}\theta_E, \sqrt{n}\theta^\perp_{E'}, \sqrt{n}V, U)|\\
		&\times g_n(\sqrt{n}(\theta_E-\theta^*_E),\sqrt{n}(\theta^\perp_{E'}-\theta^{\perp*}_{E'}),\Pi_{\sqrt{n}\theta, \sqrt{n}\theta^\perp}(\sqrt{n}V, U)).
	\end{align*}
	From the definition in \eqref{chapter3:map}, the Jacobian is a constant, hence
	\begin{equation*}
		g_n(\sqrt{n}(\bar\theta_E - \theta^*_E), \sqrt{n}(\bar\theta^\perp_{E'} - \theta^{\perp*}_{E'}), \Pi_{\sqrt{n}\bar\theta_E, \sqrt{n}\bar\theta^\perp}(\sqrt{n}\widehat V, \widehat U)).
	\end{equation*}
	By asymptotic normality and independence, for a fixed set $E$,
	\begin{equation*}
		\sqrt{n}\begin{pmatrix}
			\bar\theta_E - \theta^*_E \\
			\bar\theta^\perp_{E'} - \theta^{\perp*}_{E'} \\
			\overline\omega_n
		\end{pmatrix} \xrightarrow{d} N_{2d}\left(\begin{pmatrix}
			0_{|E|} \\
			0_{|E'|}\\
			0_d
		\end{pmatrix}, \begin{pmatrix}
			\Sigma_E & 0_{|E|,|E'|} & 0_{|E|,d} \\
			0_{|E'|,|E|} & \Sigma_{E'}^\perp & 0_{|E'|,d} \\
			0_{d,|E|} & 0_{d,|E'|} & \Omega
		\end{pmatrix}\right).
	\end{equation*}
	Replacing the pre-selection density $g_n$ in Proposition~\ref{chapter3:prop:density} with the normal density above, the proof is complete.
\end{proof}
\begin{proof}[Proof of Theorem~\ref{chapter3:theorem:conddensity}]
		Starting from the density of $(\theta_E, \theta^\perp_{E'}, V, U)$ after the change of variables and conditioning on the selection event:
		\begin{align}\label{chapter3:conddensity}
			% &\frac{\rho(\sqrt{n}\theta_E;\sqrt{n}\theta^*_E, \Sigma_E)\rho(\sqrt{n}\theta^\perp_{E'};\sqrt{n}\theta^{\perp*}_{E'}, \Sigma_{E'}^\perp)\rho(\Pi_{\sqrt{n}\theta, \sqrt{n}\theta^\perp}(\sqrt{n}V,  U); 0_d,  \Omega)\mathbbm{1}_{\mathbb{R}_+^{|E|}}(\sqrt{n}V)}
			% {\int \rho(\sqrt{n}\tilde{\theta};\sqrt{n}\theta^*_E, \Sigma_E)\rho(\sqrt{n}\theta^\perp_{E'};\sqrt{n}\theta^{\perp*}_{E'}, \Sigma_{E'}^\perp)\rho(\Pi_{\sqrt{n}\tilde{\theta}, \sqrt{n}\theta^\perp}(\sqrt{n}\tilde{V}, U); 0_d,  \Omega)\mathbbm{1}_{\mathbb{R}_+^{|E|}}(\sqrt{n}\tilde{V})d\tilde{\theta}d\tilde{V}} \nonumber \\
			\frac{\rho(\sqrt{n}\theta_E;\sqrt{n}\theta^*_E,{ \Sigma}_{E})\rho(\Pi_{\sqrt{n}\theta, \sqrt{n}\theta^\perp}(\sqrt{n}V, U); 0_d,  \Omega)\mathbbm{1}_{\mathbb{R}_+^{|E|}}(\sqrt{n}V)}
			{\int \rho(\sqrt{n}\tilde{\theta};\sqrt{n}\theta^*_E, \Sigma_E)\rho(\Pi_{\sqrt{n}\tilde{\theta}, \sqrt{n}\theta^\perp}(\sqrt{n}\tilde{V}, U); 0_d,  \Omega)\mathbbm{1}_{\mathbb{R}_+^{|E|}}(\sqrt{n}\tilde{V})d\tilde{\theta}d\tilde{V}}.
		\end{align}
		Note that the factor $\rho(\sqrt{n}\theta^\perp_{E'};\sqrt{n}\theta^{\perp*}_{E'}, \Sigma_{E'}^\perp)$ cancels out in the numerator and denominator, since it does not depend on $\theta_E$ or $V$.
		By Lemma~\ref{chapter3:density_formulation}, and following the proof of Proposition~2.4 in \citet{huang2025selective}, the numerator is proportional to
		${d}(\theta^*_E)\rho(\sqrt{n}\theta_E; L\sqrt{n}\theta^*_E+ m, Z)\rho(\sqrt{n}V; P\sqrt{n}\theta_E+ q, \Delta)$, where $c_E = 1/[(2\pi)^{|{E}|+d}|{\Sigma_E}|| \Omega|]^{1/2}$, $c(\theta^*_E) =  q^\tp \Delta^{-1} q-\frac{1}{2}(\mathcal P'(S,U)\\+\sqrt{n}\theta^\perp)^\tp \Omega^{-1}(\mathcal P'(S,U)+\sqrt{n}\theta^\perp) - \frac{1}{2}n\theta^{*\tp}_E{\Sigma_E}^{-1}\theta^*_E$ and $d(\theta^*_E) = c_E\exp\{-(L\sqrt{n}\theta^*_E+m)^\tp Z^{-1}( L\sqrt{n}\theta^*_E+ m)+c(\theta^*_E)\}$ are constants which do not depend on $\theta_E$ or $V$.
		Substitute this new form into \eqref{chapter3:conddensity} and the proof is complete.
	\end{proof}
\begin{proof}[Proof of Proposition~\ref{chapter3:map_elasticnet}]
	The subgradient of the elastic net penalty needs to be computed appropriately from the corresponding K.K.T. conditions:
	\begin{align*}
		D_p^\tp\vc&(C_n-W_n-\widehat\Theta^{-1}+\gamma\mathcal P'_{mat}(S,U)+\lambda(1-\gamma)\widehat\Theta) = 0 \\
		% &\frac{\sqrt{n}}{2}D_p^\tp \vc(C_n-\widehat\Theta^{-1})-\sqrt{n}\omega_n+\frac{\sqrt{n}}{2}\gamma D_p^\tp\vc(\mathcal P'_{mat}(S,U))+\frac{\sqrt{n}}{2}\lambda(1-\gamma) D_p^\tp\vc(\widehat\Theta) = 0 \\
		% &n^{-1/2}\grad{\widehat\Theta}-\sqrt{n}\omega_n+\gamma\mathcal P'(S,U)+\frac{\sqrt{n}}{2}\lambda(1-\gamma) D_p^\tp\vc(\widehat\Theta) = 0. \\
		% &\sqrt{n}\omega_n = n^{-1/2}\grad{\widehat\Theta} +\frac{\sqrt{n}}{2}\gamma D_p^\tp\vc(\widehat{\mathcal P}')+\frac{\sqrt{n}}{2}\lambda(1-\gamma) D_p^\tp\vc(\widehat\Theta) + o_P(1) \\
		\sqrt{n}\omega_n &= (-{J}_{n,E}{J}_{n,{EE}}^{-1}{H}_{n,{EE}})\sqrt{n}\bar\theta_E + (H_{n,E}+ \tilde C_0)\sqrt{n}\diag(S)V+\sqrt{n}\bar\theta^\perp\\ &+
		\gamma\mathcal P'(S,U) + o_P(1).
	\end{align*}	
\end{proof}
\begin{proof}[Proof of Proposition~\ref{chapter3:prop:existence_elnet}]
\tcr{For the objective function in Equation~\ref{chapter3:elnet_objective}, which we denote by $f(\Theta)$, the linear trace term is convex, the function $-\log\det(\Theta)$ is strictly convex over the cone of positive definite matrices, the $\ell_1$ norm is convex, and the squared Frobenius norm is strictly convex. Since $\lambda>0$ and $\gamma\in(0,1)$, the objective is strictly convex.
If the objective function is coercive, i.e. $f(\Theta)\rightarrow\infty$ as $\|\Theta\|_F\rightarrow\infty$ \citep{munoz2017coercive, brezis2011functional}, then a solution exists.
By the Cauchy-Schwarz inequality \citep{renaud1998some},
\begin{align*}
	|\tr((C_n-W_n)^\tp\Theta)|^2 \le \tr((C_n-W_n)^\tp(C_n-W_n))\tr(\Theta^\tp\Theta) = \|C_n-W_n\|_F^2\|\Theta\|_F^2 \\
	% |\tr((C_n-W_n)\Theta)| \le \|C_n-W_n\|_F\|\Theta\|_F \\
	\tr((C_n-W_n)\Theta) \ge -\|C_n-W_n\|_F\|\Theta\|_F\\
	f(\Theta) \ge \frac{\lambda(1-\gamma)}{2}\|\Theta\|_F^2 - \|C_n-W_n\|_F\|\Theta\|_F - \log\det(\Theta) + \lambda\gamma\|\Theta\|_1.
\end{align*}
The quadratic term dominates the linear and logarithmic terms as $\|\Theta\|_F\rightarrow\infty$, implying that $f(\Theta)\rightarrow\infty$, proving that the objective function is coercive. Since the objective function is strictly convex and coercive, it has a unique minimizer.
}
\end{proof}
\begin{proof}[Proof of Theorem~\ref{chapter3:theorem:mle}]
The results follow from the arguments in the proof of Theorem~4.1 in \citet{huang2025selective}. 
Using the reparameterization $\eta_E = Z^{-1}(L\sqrt{n}\theta^*_E+ m)$ and the fact that the Jacobian term is constant by Lemma~\ref{chapter3:prop:density}, we analytically maximize the selective log-likelihood. This yields the stated expression for the selective MLE.
Taking the Hessian of the negative selective log-likelihood and evaluating it at the selective MLE yields the expression for the observed information matrix.
\end{proof}

\section{Coverage in the high-dimensional case}

We present results for a high-dimensional case, similar to the one considered in the application, with $n=60$ observations generated from a $p=70$ dimensional multivariate normal distribution.
Table~\ref{chapter3:tab:coverage_lasso} shows the coverage results for the different methods.

\begin{table}[ht]
\centering
\caption{Simulated coverage for the graphical lasso penalty with $n=60$ and $p=70$, averaged over 500 runs. The nominal coverage is 95\%. Prec.: precision matrix entriess; P. Corr.: partial correlations; NS: node strength; BS: bridge strength.}\label{chapter3:tab:coverage_lasso}
\begin{tabular}{rcccccc}
 & \textbf{Prec}. & \textbf{P. Corr.} & \textbf{NS} & \textbf{NS Diff.} & \textbf{BS} & \textbf{BS Diff.} \\ 
  \midrule
  DS (10\%) & 0.954 & 0.930 & 0.934 & 0.970 & 0.936 & 0.974 \\ 
  DS (50\%) & 0.958 & 0.919 & 0.932 & 0.962 & 0.929 & 0.967 \\ 
  DS (90\%) & 0.992 & 0.769 & 0.765 & 0.836 & 0.774 & 0.841 \\ 
  POLY & 0.962 & -- & -- & -- & -- & -- \\ 
  NAIVE & 0.677 & 0.688 & 0.451 & 0.800 & 0.591 & 0.812 \\ 
  SMLE (0.5) & 0.980 & 0.974 & 0.959 & 0.989 & 0.959 & 0.992 \\ 
  SMLE (0.63) & 0.972 & 0.957 & 0.966 & 0.980 & 0.961 & 0.985 \\ 
  SMLE (0.79) & 0.960 & 0.939 & 0.953 & 0.971 & 0.950 & 0.975 \\ 
  SMLE (0.99) & 0.949 & 0.923 & 0.932 & 0.956 & 0.934 & 0.967 \\ 
  SMLE (1.25) & 0.936 & 0.909 & 0.915 & 0.954 & 0.921 & 0.962 \\ 
\end{tabular}
\end{table}

As expected, the naive method presents under-coverage for all targets, due to not accounting for selection. Data splitting with 90\% of the sample used for selection also presents under-coverage, as well as a 6\% rate of computational failure, which is likely due to the small sample size used for inference. 
The polyhedral method without randomization shows satisfactory coverage for the precision matrix elements, but is not applicable for the other targets. Our SMLE method with randomization achieves the desired coverage across all targets and values of the randomization standard deviation $\sigma_\omega$, both for the graphical lasso and elastic net penalties. Data splitting with 10\% or 50\% of the sample used for selection also attains the nominal coverage. We further investigate a comparison with these methods.

\section{Simulation results for the elastic net penalty}

\begin{figure}[ht]
	\centering
	\includegraphics[width=0.45\linewidth]{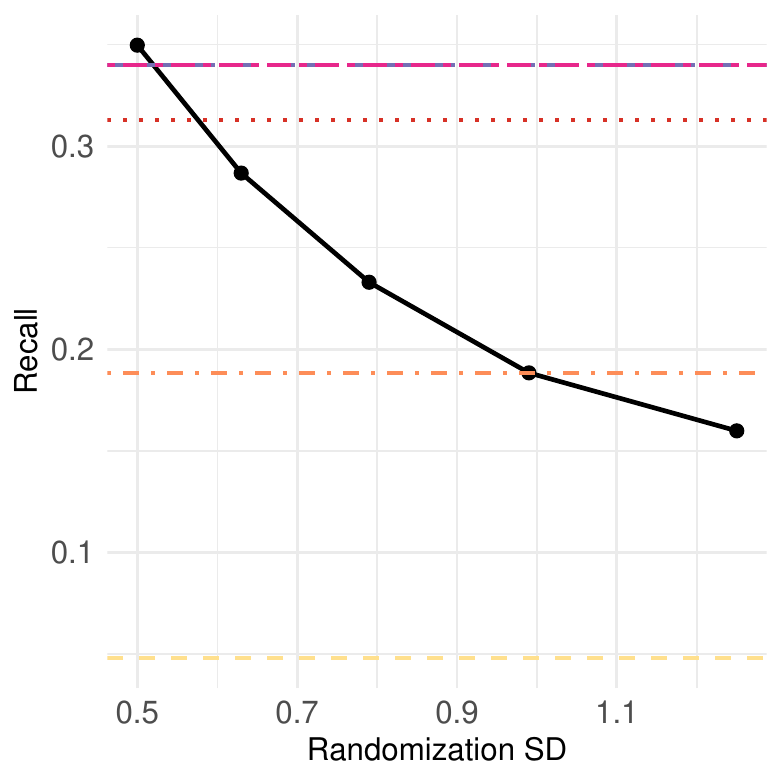}
	\includegraphics[width=0.45\linewidth]{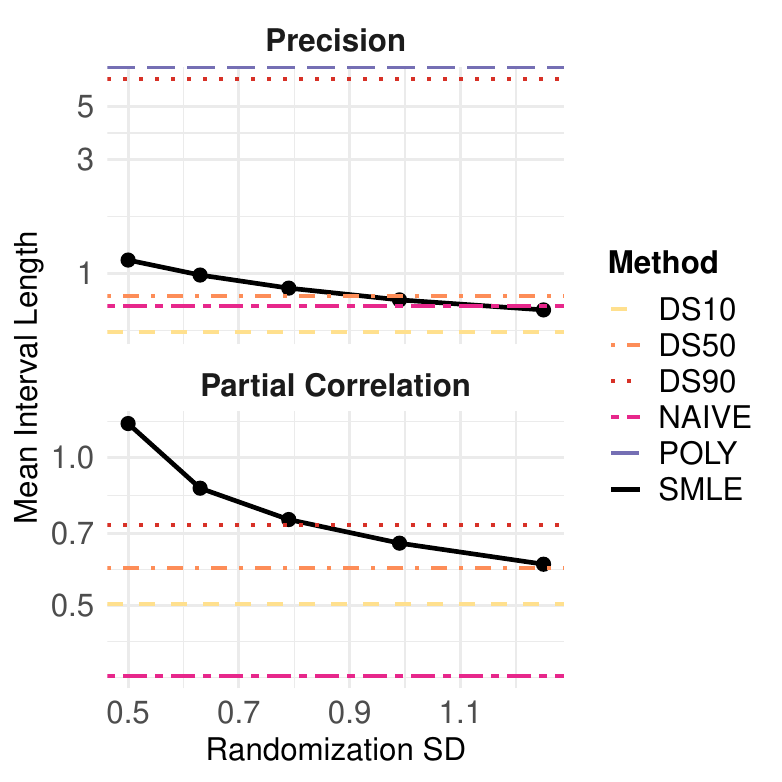}
	\caption{Left: Recall of the selection step for randomized graphical elastic net, as the amount of randomization varies, compared to the other methods. Right: Average length of the confidence intervals for the precision and partial correlation elements for randomized graphical elastic net as the amount of randomization varies, compared to the other methods. Each point is the average over 500 runs, with $n = 60$ and $p=70$.}
	\label{chapter3:fig:recall_elnet_sd}
\end{figure}

We present additional simulation results for the randomized graphical elastic net in Table~\ref{chapter3:tab:coverage_elnet} and Figure~\ref{chapter3:fig:recall_elnet_sd}. 
Similar considerations as in Section~\ref{chapter3:sec:simulations} apply here. The selective inference methods (data splitting with different amounts of randomization and the proposed selective MLE with different amounts of randomization) achieve the desired nominal coverage of 95\% for all the considered quantities. The polyhedral method also attains the nominal coverage, while the naive method presents severe undercoverage, as well as data splitting with 90\% of the data used for selection.
As the amount of randomization increases, the recall of the selection step decreases, while the average length of the confidence intervals decreases. 
Note that the 90\% data splitting also has a 7.4\% rate of computational failure, as well as the selective MLE with 0.5 standard deviation of the randomization (1.6\% failure rate). These failures are due to numerical issues \tcr{in the inference step}, and depend on the density of the true graph, as well as the penalty parameter. Further investigation of these numerical issues is left for future work.

\begin{table}[ht]
\centering
\caption{Simulated coverage for the graphical elastic net penalty with $n=60$ and $p=70$, averaged over 500 runs. The nominal coverage is 95\%. Prec.: precision matrix entries; P. Corr.: partial correlations; NS: node strength; BS: bridge strength.}\label{chapter3:tab:coverage_elnet}
\begin{tabular}{rcccccc}
  & \textbf{Prec.} & \textbf{P. Corr.} & \textbf{NS} & \textbf{NS Diff.} & \textbf{BS} & \textbf{BS Diff.} \\ 
  \midrule
DS (10\%) & 0.953 & 0.930 & 0.934 & 0.970 & 0.936 & 0.974 \\ 
  D (50\%) & 0.958 & 0.919 & 0.932 & 0.962 & 0.929 & 0.967 \\ 
  D (90\%) & 0.992 & 0.766 & 0.763 & 0.835 & 0.772 & 0.841 \\ 
  POLY & 0.960 &  &  &  &  &  \\ 
  NAIVE & 0.685 & 0.694 & 0.459 & 0.806 & 0.597 & 0.816 \\ 
  SMLE (0.5) & 0.918 & 0.940 & 0.967 & 0.985 & 0.965 & 0.988 \\ 
  SMLE (0.63) & 0.928 & 0.932 & 0.951 & 0.971 & 0.949 & 0.977 \\ 
  SMLE (0.79) & 0.937 & 0.933 & 0.946 & 0.966 & 0.947 & 0.977 \\ 
  SMLE (0.99) & 0.943 & 0.934 & 0.947 & 0.963 & 0.945 & 0.973 \\ 
  SMLE (1.25) & 0.944 & 0.930 & 0.942 & 0.964 & 0.942 & 0.972 \\ 
\end{tabular}
\end{table}

\section{Extra tables}

\begin{table}[H]
\centering
\caption{Post-selection inferential targets in the learned graph. For bridge-related quantities, $\mathcal{B}(j)$ denotes the set of nodes belonging to communities different from that of node $j$. The centrality metrics can also be taken with respect to the partial correlations.}
\label{tab:targets}
\begin{tabular}{lll}
\textbf{Target} & \textbf{Definition} & \textbf{Inference} \\
\midrule
Edge parameter
& $T_{ij}(\theta)=\theta_k,\quad k=k(i,j)\in E$
& Selective MLE \\

Exp. influence
& $T_{\mathrm{EI}}^{(j)}(\theta)=\displaystyle\sum_{k\in E:\,k\leftrightarrow(i,j)} \theta_k$
& Selective MLE \\

Partial correlation
& $T_{ij}(\theta)= -\theta_k\big/ \sqrt{\theta_{k(ii)}\,\theta_{k(jj)}}$
& Selective delta method \\

Node strength
& $T_{\mathrm{NS}}^{(j)}(\theta)=\displaystyle\sum_{k\in E:\,k\leftrightarrow(i,j)} |\theta_k|$
& Selective bootstrap \\

Diff. in node strength
& $T_{\mathrm{NS}}^{(j_1,j_2)}(\theta)
   =T_{\mathrm{NS}}^{(j_1)}(\theta)-T_{\mathrm{NS}}^{(j_2)}(\theta)$
& Selective bootstrap \\

Bridge strength
& $T_{\mathrm{BS}}^{(j)}(\theta)
   =\displaystyle\sum_{k\in E:\,i\in\mathcal{B}(j)} |\theta_k|$
& Selective bootstrap \\

Diff. in bridge strength
& $T_{\mathrm{BS}}^{(j_1,j_2)}(\theta)
   =T_{\mathrm{BS}}^{(j_1)}(\theta)-T_{\mathrm{BS}}^{(j_2)}(\theta)$
& Selective bootstrap \\

Bridge exp. influence
& $T_{\mathrm{BEI}}^{(j)}(\theta)
   =\displaystyle\sum_{k\in E:\,i\in\mathcal{B}(j)} \theta_k$
& Selective MLE \\
\end{tabular}
\end{table}

\begin{algorithm}
	\caption{Obtain the selective maximum likelihood estimate (SMLE) and Fisher information matrix. The computation of the necessary quantities is detailed in Table~\ref{chapter3:tab:quantities}.}\label{chapter3:algorithm1}
	\begin{algorithmic}[1]
	\setlength{\baselineskip}{1.2\baselineskip}
		\Require Scaled, centered data matrix $X$ of size $n \times p$
		\Require Fixed penalty $\lambda=O_P(n^{-1/2})$
		\Require Randomization variance matrix $\Omega$ of size $d \times d$, $d=p(p+1)/2$
		\Ensure Selective m.l.e. $\widehat\theta^{SMLE}_E$ and asymptotic variance $\widehat\Sigma_E^{SMLE}$
		% Selection subpart
		\Statex \textbf{Selection:}
		\State Generate $\omega\sim N_d(0_d, \Omega)$
		\State $W_n \gets \vh^{-1}(\omega/\sqrt{n})$
		\State $\diag(W_n) \gets 2\diag(W_n)$.
		\State $C_n \gets X^\tp X/n$
		\State $\widehat\Theta \gets \Call{GraphicalLasso}{C_n-W_n, \lambda}$
		% Inference subpart
		\Statex \textbf{Inference:}
		\State $\bar\Theta \gets \Call{ConstrainedMLE}{C_n, \bar\Theta_{E'}=0}$
		\For{each $j$ from 1 to $d$}
		\State $\text{subgrad}_j \gets -\ell'_j+\omega_{n,j}$ \Comment{Penalty (sub)gradient from KKT conditions}
		\EndFor
		\State Compute the quantities in Table~\ref{chapter3:tab:quantities}
		\State $\widehat V^{SMLE} \gets \argmin_b \{b^\tp\Delta^{-1}b/2-b^\tp\Delta^{-1}(P(\sqrt{n}\bar\theta_E)+\sqrt{n}q)-\sum_{j=1}^{|E|}\log(b_j)\}$
		\State $\widehat\theta^{SMLE}_E \gets L^{-1}\bar\theta_E+L^{-1}ZP^\tp \Delta^{-1}(P\bar\theta_E+q-\widehat V^{SMLE}/\sqrt{n})-L^{-1}m$
		\State $\widehat\Sigma_E^{SMLE} \gets [L^{-1}ZL^{-\tp}+L^{-1}Z[P^\tp\Delta^{-1}P$
		\Statex \hspace{50pt} $-P^\tp\Delta^{-1}(\Delta^{-1}+\diag(1/\widehat V^{SMLE})^2)^{-1}\Delta^{-1}P]ZL^{-\tp}]^{-1}$
	\end{algorithmic}
\end{algorithm}

\begin{table}[H]
	\centering
	\begin{tabular}{l l}
		Quantity & Value \\
		\midrule
		$\omega_n$ & $\omega/\sqrt{n}$ \\
		$D_p$  & duplication matrix of order $p$ \\
		$H_n$ & $D_p^\tp(\bar\Sigma\otimes\bar\Sigma)D_p$ \\
		$J_n$ & $\sum_{h=1}^n[D_p^\tp\vc(-\bar\Sigma+X_hX_h^\tp) \vc(-\bar\Sigma+X_hX_h^\tp)^\tp D_p]/4n$ \\
		$E$ & $\{j:\vh(\widehat\Theta)_j\neq 0\}$ \\
		$E'$ & $\{j:\vh(\widehat\Theta)_j= 0\}$ \\
		$\Sigma_E$ & $H_{n,EE}^{-1}J_{n,EE}H_{n,EE}^{-1}$ \\
		$\bar\theta$ & $\vh(\bar\Theta)$ \\
		$\bar\Sigma$ & $\bar\Theta^{-1}$ \\
		$A_E$ & $H_{n,E'E}-J_{n,E'E}J_{n,EE}^{-1}H_{n,EE}$ \\
		$\theta^\perp_{E'}$ & $D_p^\tp\vc(-\bar\Sigma+C_n)/2-A_E\bar\theta_E$ \\
		$\theta^\perp_E$ & $0_{|E|}$ \\
		$C_1$ & $-J_{n,E}J_{n,EE}^{-1}H_{n,EE}$ \\
		$C_2$ & $ H_{n,E}\diag(\mbox{sign}(\vh(\widehat\Theta)_E))$ \\
		$\ell'$ & $D_p^\tp\vc(-\widehat\Sigma+C_n)/2$ \\
		$\Delta$ & $(C_2^\tp \Omega^{-1} C_2)^{-1}$ \\
		$P$ & $-\Delta C_2^\tp \Omega^{-1} C_1$ \\
		$q$ & $-\Delta C_2^\tp \Omega^{-1} (\mbox{subgrad}+\theta^\perp)$ \\
		$Z$ & $(\Sigma_E^{-1}-P^\tp \Delta^{-1}P+C_1^\tp \Omega^{-1}(\mbox{subgrad}+\theta^\perp))^{-1}$ \\
		$L$ & $Z\Sigma_E^{-1}$ \\
		$m$ & $ZP^\tp\Delta^{-1}q-C_1^\tp\Omega^{-1}(\mbox{subgrad}+\theta^\perp)$ \\
	\end{tabular}
	\caption{Quantities for Algorithm~\ref{chapter3:algorithm1}}
	\label{chapter3:tab:quantities}
\end{table}
\end{appendices}

\end{document}